\documentstyle[graphicx,preprint,aps]{revtex}  

\begin{document}

\tightenlines
\draft
\title{Relativistic many-body calculations of energy levels, 
       hyperfine constants, electric-dipole matrix elements and static
       polarizabilities for alkali-metal atoms.}

\author{M. S. Safronova, W. R. Johnson, and A. Derevianko}
\address{Department of Physics, Notre Dame University, \\
         Notre Dame, IN 46556}
\date{\today}
\maketitle
\begin{abstract}

Removal energies and hyperfine constants
of the lowest four $ns$, $np_{1/2}$ and $np_{3/2}$ states in Na, 
K, Rb and Cs are calculated; removal energies 
of the $n$=7--10 states and hyperfine constants 
of the $n$=7 and 8 states in Fr are also calculated. 
The calculations are based on the relativistic single-double (SD)
approximation in which single and double excitations of 
Dirac-Hartree-Fock (DHF) wave functions are included to all-orders in perturbation theory.  
Using SD wave functions, accurate values of removal energies, electric-dipole matrix elements 
and static polarizabilities are obtained, however, SD wave functions give
poor values of magnetic-dipole hyperfine constants for heavy atoms.
To obtain accurate values of hyperfine constants for heavy atoms,    
we include triple excitations partially in the wave functions. 
The present calculations provide the basis for reevaluating PNC amplitudes in Cs and Fr.
\end{abstract}

\pacs{31.15.Ar, 31.25.Jf, 32.10.Fn, 32.10.Dk, 32.70.Cs}
%\twocolumn
\section{Introduction}
\label{Intro}
 
Energy levels, transition matrix elements, hyperfine constants, and static polarizabilities
for low-lying $s_{1/2}$, $p_{1/2}$ and $p_{3/2}$ states in alkali-metal atoms are studied 
systematically using the relativistic single-double (SD) method in which 
single and double excitations of the Dirac-Hartree-Fock (DHF)
wave function are included to all orders of perturbation theory.  
The SD method was applied previously to study properties of Li and Be$^+$ \cite{liao}, 
Li, Na, and Cs \cite{liu}, Cs \cite{csao}, 
and Na-like ions with $Z$ ranging from 11 to 16 \cite{Na}.
In the latter study,
the theoretical removal energies for Na-like ions, when corrected for the Lamb shift, 
agreed with experiment at the 1--20~cm$^{-1}$ level of accuracy for all states considered, while 
theoretical hyperfine constants and dipole matrix elements typically agreed with precise 
measurements to better than 0.3\%.

Energies of alkali-metal atoms have been calculated 
to high precision in \cite{EKI} using the relativistic coupled-cluster (CC) method; however,
there have been no systematic CC studies of hyperfine constants or transition amplitudes
for alkali-metal atoms. All-order methods are needed for such studies  
since correlation corrections are large in alkalis and
low-order many-body perturbation theory (MBPT) does not give accurate results 
for atomic properties.  
In K, Rb, and Cs, third-order MBPT gives ground state energies in poorer agreement 
with experiment than second-order MBPT, as illustrated in Table~\ref{tab1} where
zeroth-order DHF energies are tabulated together with 
second- and third-order MBPT corrections.
The differences $\Delta^{(k)}$, $k=0,2,3$ between  experimental energies 
and accumulated MBPT values shown in  
Table~\ref{tab1} oscillate above and below the experimental values
and show no sign of convergence.
In the SD approximation, an important subset of MBPT diagrams is iterated to all orders
in perturbation theory, leading to energies in excellent agreement with experiment.

During the last few years, lifetimes of the lowest $p_{1/2}$ and $p_{3/2}$ 
levels have been measured to high precision
for all alkali-metal atoms \cite{J,V,W,S,oT,Y,nT}, 
yielding experimental dipole matrix elements accurate to 0.1\%--0.25\%.
In the present work, electric-dipole matrix elements for 
$n'p-ns$ transitions in alkalis from Na to Fr are evaluated for $n=N, N+1$ and $n'=N, \ldots, N+3$,
where $N$ is the principle quantum number of the ground state. 
Matrix elements and energies from the present SD calculation were 
used in Ref.~\cite{new} to study polarizabilities of alkali-metal atoms.

In this paper, we discuss our calculation of static polarizabilities
in detail and show that the {\em ab-initio} SD results 
are in excellent agreement with the values recommended in \cite{new}.
We also calculate Stark-induced scalar and vector transition 
polarizabilities for $Ns-(N+1)s$ transitions.
The accuracy of our calculations is discussed and 
recommended values of scalar and vector polarizabilities are provided for Cs and Fr.
These values are needed for the interpretation of experiments on parity 
nonconservation in atoms \cite{Wieman}.
 
   A systematic study of hyperfine constants for
$s$, $p_{1/2}$ and $p_{3/2}$ levels is also presented. 
The accuracy of SD calculations of hyperfine 
constants for alkali-metal atoms decreases rapidly from 0.3\% for Na to 7\% for Cs. 
To obtain more accurate values for heavy alkalis, 
it was found necessary to include triple excitations to the wave functions partially. 
The derivation of an approximate single-double partial triple
(SDpT) wave function is given in the following section.

In summary, we study low-lying $s$ and $p$ levels in alkali-metal atoms in the relativistic
SD and SDpT approximations and find excellent agreement with other high-precision calculations
and with available experimental data. The SDpT wave functions from the present calculations
will be used later to evaluate parity nonconserving (PNC) amplitudes
in cesium and francium. 

\section{Triple Excitations}
\label{Method}
The all-order single-double method was described previously in
Refs.~\cite{liao,liu,csao,Na}.
Briefly, we represent the wave function $\Psi_v$ of  
a one valence electron atom as  $\Psi_v \approx \Psi_v^{\rm SD}$ with
\begin{eqnarray}
\lefteqn{\Psi_v^{\rm SD} =  \left[ 1 + \sum_{ma} \, 
\rho_{ma} a^\dagger_m a_a 
 +  \frac{1}{2} \sum_{mnab} \rho_{mnab} a^\dagger_m a^\dagger_n a_b a_a 
 \right. } \hspace{0.5in} \nonumber\\
&& \left. +\sum_{m \neq v} \rho_{mv} a^\dagger_m a_v + 
\sum_{mna} \rho_{mnva} a^\dagger_m a^\dagger_n a_a a_v 
\right]  \Phi_v  ,
\label{eqf}
\end{eqnarray}
where $\Phi_v$ is the lowest-order atomic state function, 
which is taken to be the {\em frozen-core} DHF wave function
of a state $v$.
In this equation, $a^\dagger_i$ and $a_i$ are creation and annihilation operators,
respectively, for state $i$. Indices at the beginning
of the alphabet, $a$, $b$, $\cdots$, refer to occupied core states, those in
the middle of the alphabet $m$, $n$, $\cdots$, refer to excited states, and
$v$ refers to valence orbital.
Substituting the wave function (\ref{eqf}) into the many-body
Schr\"{o}dinger equation, where the Hamiltonian is taken to be the relativistic {\em no-pair} 
Hamiltonian with Coulomb interactions \cite{nopair}, one obtains the coupled 
equations for single- and double-excitation coefficients $\rho_{mv}$, $\rho_{ma}$, $\rho_{mnva}$,
and $\rho_{mnab}$. The coupled equations are solved iteratively for the 
excitation coefficients. We use the resulting SD wave functions to evaluate hyperfine constants 
and electric-dipole matrix elements. 
%\onecolumn
  It was shown in \cite{Na} that the SD energies are not
complete in third order and that the missing third-order energy 
contributions are associated with the omitted triple-excitations.
In \cite{Na}, we calculated the missing third-order terms
separately and added them to the final energies. 
It can be shown that one-body matrix elements calculated using SD wave functions 
are complete through third order. 
As mentioned in the introduction, hyperfine 
constants for heavy alkali-metal atoms 
are not determined to high precision in the SD 
approximation.  However, by adding the two triple-excitation terms 
$\rho_{mnrvab}$ and $\rho_{mnrabc}$ to the SD wave function,
we automatically include the missing third-order energy and also substantially
improve the accuracy of our calculations of hyperfine constants. 
The corrected wave function is 
\begin{eqnarray}
\lefteqn{  \Psi_v \approx \Psi_v^{\rm SD}
 + \left[ \frac{1} {6}\sum_{mnrab} \rho_{mnrvab} 
 a_m^{\dagger } a_n^{\dagger} a_r^{\dagger } a_b a_aa_v \right.}\hspace{1.5in}\nonumber \\
&& + \left. \frac{1} {18}\sum_{mnrabc} \rho_{mnrabc} 
a_m^{\dagger } a_n^{\dagger} a_r^{\dagger } a_c a_b a_a\right] \Phi_v \, ,
\label{SDT}
\end{eqnarray}
where  $\Psi_v^{\rm SD}$ is single-double wave function in Eq.~(\ref{eqf}).
The addition of the core term $\rho_{mnrabc}$ is necessary to preserve the symmetry
relation
$\rho_{mnva}=\rho_{nmav}$. Carrying out the calculations, we obtain the following
equations for the energy, single- and double-excitation coefficients:
\begin{eqnarray}
\lefteqn{\delta E_v = {\rm (SD)}
+\sum_{mnab}g_{abmn} {\rho }_{mnvvab}\; , } \hspace{0.5in} \label{asd} \\
\lefteqn{\left( \varepsilon _a-\varepsilon _m \right) \rho_{ma} 
= {\rm (SD)}+\sum_{nrbc} g_{bcnr} \rho_{mnrabc}\; ,} \hspace{0.5in}\\ 
\lefteqn{\left( \varepsilon _v-\varepsilon _m +\delta E_v\right) \rho_{mv}
={\rm (SD)}+\sum_{nrab}g_{abnr}\rho_{mnrvab}\; ,} \hspace{0.5in} \label{bsd}\\ 
\lefteqn{
 \left( \varepsilon _a+\varepsilon _b-\varepsilon _m-\varepsilon_n \right) \rho _{mnab}
= {\rm (SD)} }\hspace{0.5in} \nonumber\\
&& - \sum_{rcd} g_{cdar} \rho_{mnrbdc}-\sum_{rcd} g_{cdbr} \rho_{nmradc} 
   - \sum_{rsc}g_{cmrs} \rho_{snrbac} -\sum_{rsc}g_{cnrs} \rho_{smrabc}\; , \\
\lefteqn{ \left( \varepsilon _a+\varepsilon _v-\varepsilon _m-\varepsilon _n 
   +\delta E_v\right) \rho _{mnva}={\rm (SD)}}\hspace{0.5in}\nonumber\\
&&+\sum_{rcb}g_{bcar} \rho_{mnrvcb}
+\sum_{rc} g_{bcvr} \rho_{mnrabc}
+\sum_{rsb}g_{bmrs} \rho_{srnvba} 
+\sum_{rsb}g_{bnrs} \rho_{srmvab} \;.
\end{eqnarray}
In the above equations, we write out only those terms arising from the triple
excitations. The quantities $\varepsilon_i$ are single-particle energies and 
$\delta E_v$ is the correlation correction to the valence energy. Below, we use the
notations $\varepsilon _{mn} = \varepsilon_m+\varepsilon_n$, $\widetilde{g}_{abcd} = g_{abcd} 
-g_{abdc}$ and $\widetilde{\rho}_{abcd} = \rho_{abcd} -\rho_{abdc}$.
The contributions from the single- and double-excitation coefficients,
designated by (SD) above, are given in Ref.~\cite{Na}.
We require that the triple-excitation coefficients $\rho_{mnrvab}$
and $\rho_{mnrabc}$ be antisymmetric with respect to any non-cyclic 
permutation of the indices  $mnr$, $vab$ and $mnr$, $abc$,
respectively, and we obtain the following equations for the triple-excitation 
coefficients:
\begin{eqnarray}
\lefteqn{
 \left(\varepsilon_a+\varepsilon_b+\varepsilon_c-\varepsilon_m-
  \varepsilon_n-\varepsilon_r \right) \rho _{mnrabc}=}\hspace{0.5in}\nonumber\\
&&
\sum_{\scriptsize
\begin{array} {ccc}
123&=&\{mnr\} \\ 
1'2'3'&=&\{abc\} 
\end{array}}
 \frac12 \left(\frac12 g_{121'2'} \rho_{33'} 
 -\sum_d g_{1d1'2'} \rho_{23d3'} +\sum_s g_{23s3'} \rho_{1s1'2'} \right)+
[ {\rm triples} ] \;, \label{eq8}
\end{eqnarray}
\begin{eqnarray}
\lefteqn{
 \left(\varepsilon_a+\varepsilon_b+\varepsilon_v-\varepsilon_m- 
 \varepsilon_n-\varepsilon_r +\delta E_v \right) \rho _{mnrvab}=}\hspace{0.5in}\nonumber\\
&&
\sum_{\scriptsize
\begin{array} {ccc}
123&=&\{mnr\} \\ 
1'2'3' &=&\{vab\} 
\end{array}}
 \frac12 \left(\frac12 g_{121'2'} \rho_{33'} -
 \sum_c g_{1c1'2'} \rho_{23c3'} +\sum_s g_{23s3'} \rho_{1s1'2'} \right)+
[{\rm triples} ]\;, \label{eq9}
\end{eqnarray}
where [{\rm triples}] groups together terms containing $\rho _{mnrvab}$ or $\rho _{mnrabc}$.
In the above equations, the notation $123 = \{mnr\}$ designates symbolically that the indices 
$123$ range over all six permutations of the indices $mnr$; even permutations contribute with
a positive sign while odd permutations contribute with a negative sign.
The relatively small contributions from single- and triple-excitations 
on the right-hand sides of
Eqs.~(\ref{eq8}) and (\ref{eq9}) are omitted in the present study.

The dominant triples corrections arise from the triple contributions to $\delta E_v$ 
and $\rho_{mv}$  given in  Eq.~(\ref{asd}) and Eq.~(\ref{bsd}), respectively.
Solving the equation for $\rho_{mnrvab}$ and substituting the resulting expression 
into Eq.~(\ref{asd}), we find
\begin{eqnarray}
\lefteqn{
\delta E_{v} \approx {\rm (SD)} + \sum_{mnab}
\frac {\widetilde{g}_{abmn}}{\varepsilon_{ab}-\varepsilon_{mn}}
\left\{   \sum_c \widetilde{g}_{cmav} \widetilde{\rho}_{nvbc} 
   +      \sum_s \widetilde{g}_{nvas} \widetilde{\rho}_{msvb} 
   + \sum_c \widetilde{g}_{cvbv} {\rho}_{mnca} 
   \right. }\hspace{0.25in} \nonumber\\
&& \left. 
   + \sum_s \widetilde{g}_{mvsv} {\rho}_{nsba}
   + \sum_c {g}_{cmab} \widetilde{\rho}_{vnvc} 
   + \sum_s {g}_{mnas} \widetilde{\rho}_{vsvb}
   + \sum_s  {g}_{mnvs} \rho_{vsba} 
   + \sum_c  {g}_{cvba} \rho_{mnvc} \right\}\, . \label{ceq}
\end{eqnarray} 
 Repeating these steps for $\rho_{mv}$, we obtain from Eq.~(\ref{bsd}):
\begin{eqnarray}
\lefteqn{
\left(\varepsilon _v-\varepsilon _m + \delta E_v\right) \rho_{mv}\approx {\rm (SD)}
-\sum_{nrab} \frac{\widetilde{g}_{abnr}} {\left(\varepsilon_{ab}-\varepsilon_{nr} \right)}
 \left\{ 
  \sum_{c} \widetilde{g}_{ncva} \widetilde{\rho}_{rmcb}
 -\sum_{s} \widetilde{g}_{rmsa} \widetilde{\rho}_{snvb}
 +\sum_{c} \widetilde{g}_{mcva} \rho_{nrcb}
 \right. }\hspace{0.35in} \nonumber\\
 && \left.
 -\sum_{s} \widetilde{g}_{rmsv} \rho_{nsab}
 +\sum_{c} g_{ncab} \widetilde{\rho}_{rmcv} 
 -\sum_{s}g_{nrsa} \widetilde{\rho}_{smvb}   
 -\sum_{s} g_{nrsv} \rho_{msab}
 + \sum_{c} g_{mcab} \rho_{nrcv}
 \right\} \, .\label{deq}
\end{eqnarray}
In our numerical studies, we use the approximation 
\[
\frac {\widetilde{g}_{abmn}}{\varepsilon_{ab}-\varepsilon_{mn}} \approx \widetilde{\rho}_{mnab} \, 
\]
in Eqs.~(\ref{ceq}) and (\ref{deq}).
In the present calculations, we include triples in the $\rho_{mv}$ and  $\delta E_v $ 
equations only. 
As discussed above, only double-excitation terms are considered in the equations for the
triple-excitation coefficients. Finally, {\em explicit} 
triple-excitation corrections to matrix elements
are omitted; only indirect corrections caused by modification 
of $\delta E_v$ and $\rho_{mv}$ are included.  The modified matrix elements 
are evaluated as described in Ref.~\cite{Na}. 
In the approximation used here, all third-order corrections to $\delta E_v$ are 
automatically included.  

\section{Results and Discussions}
\subsection{Removal energies and fine structure}

The SD equations are set up in a finite basis and solved iteratively 
to give the single- and double-excitation coefficients $\rho_{ma}$, $\rho_{mv}$, 
$\rho_{mnab}$ and $\rho_{mnva}$, and the correlation energy $\delta E_v$. 
The basis orbitals used to define the single-particle states are 
linear combinations of B-splines \cite{31}. 
For each angular momentum state, the basis set consisted of 40 
basis orbitals constructed from 40 B-splines of order 7. 
In our iterative calculations, we used only 35 of the 40 orbitals. 
The B-spline basis orbitals were interpolated onto a 250 point 
nonlinear radial grid. 
All orbitals were constrained to a large spherical cavity; the cavity radii were 
chosen to be 110 a.u.\ for Na, 100 a.u.\ for K and Rb, 75 a.u.\ for Cs, 
and 90 a.u.\ for Fr. Such large cavities were needed to accommodate the highly
excited states considered here. 
The DHF energies of the lowest 3-4 $s$ and $p$ states were reproduced to
five or more significant digits by the B-spline basis functions.  
Generally, the larger values of $n$ had lower accuracy, which is unimportant 
owing to the decreasing size of correlation corrections with increasing $n$. 
Terms in the angular-momentum decomposition with angular momentum $l$ from 0 to 6 were retained 
in the basis and the partial-wave contributions were extrapolated
to give the final values of the correlation energy. The extrapolation
procedure is described in \cite{Na}. 
For the case of Fr, only partial waves with $l \leq 5$ were retained 
because of computational limitations, and the extrapolation procedure was simplified,
leading to somewhat lower accuracy.

Contributions to the energy from the Breit interaction (with all-order correlation corrections)
were obtained as expectation values of the Breit operator 
using SD wave functions, as described in \cite{Na}.
Breit corrections were found to be less than 
15cm$^{-1}$ in all cases.

\begin{figure}
\centerline{\includegraphics*[scale=0.667]{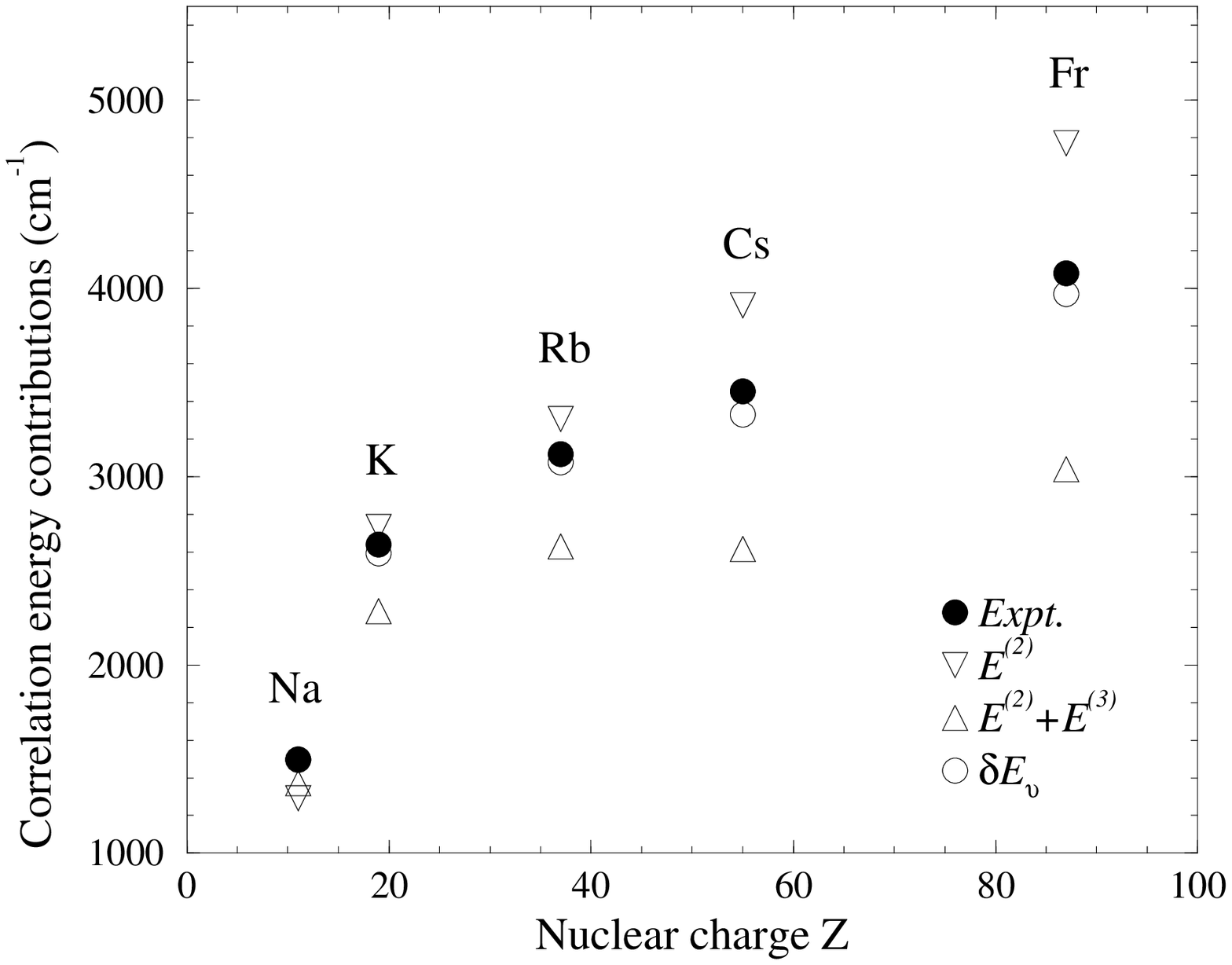}}
\caption{Comparisons of MBPT and SD correlation corrections to ground state energies
for alkali-metal atoms.
\label{fig1} }    
\end{figure}

In Fig.~\ref{fig1}, values of
$\delta E_v$ for the ground states of the alkalis, corrected for missing triples, are compared with 
the second-order energy $E^{(2)}$, the third-order energy $E^{(2)}$+$E^{(3)}$, and with the 
experimental correlation energy {\it Expt.}
Differences between $\delta E_v$ and $E^{(2)}$+$E^{(3)}$ are from fourth- and higher-order
terms in the MBPT expansion.  It is clear from the figure that the SD procedure resolves
the problem of poor convergence of MBPT discussed previously and shown in Table~\ref{tab1}, and that the SD
ground-state correlation energies are in good agreement with experimental values.
Contributions from fourth- and higher-order corrections increase 
from 8\% of the total correlation energy for Na to 24\% for Fr.  
Differences with measurements
for the ground-state correlation energy range from 0.1\% for Na to 2.7\% for Fr.
The SD approximation, therefore, accounts for a dominant fraction of 
the fourth- and  higher-order correlation energy.
Correlation corrections for lowest $p_{1/2}$ states are about 3 times smaller than those for the ground states.
The relative contributions of higher-order corrections are found to be approximately 
the same for $ns$ and $np$ states.

A detailed comparison of removal energies for $s$ and $p_{1/2}$ states with experiment is given in 
Table \ref{tab2}. The experimental data used in this comparison are from 
Ref.~\cite{NIST} except for Fr, 
where experimental energies compiled in \cite{Fr} 
and results of recent measurements \cite{8s9s} are used.
For Na, our theoretical uncertainty (from extrapolation) ranges from 0.4~cm$^{-1}$ 
for the $3s$ state to 0.04~cm$^{-1}$ for the $6s$ state;
this uncertainty increases for heavier alkalis.
The agreement with experiment is excellent for Na, 
where the $6s$ energy differs from experiment by 0.14~cm$^{-1}$
and the $3s$ energy differs by 2~cm$^{-1}$. 
The corresponding differences are 5--48~cm$^{-1}$ in K,
7--42~cm$^{-1}$ in Rb,
17--145~cm$^{-1}$ in Cs, and 16--114~cm$^{-1}$ in Fr.
Agreement with experiment improves substantially with $n$ since correlation corrections decrease.
Our results for removal energies of $np_{1/2}$ states are in excellent agreement 
with experiment for all states considered.
For $np_{1/2}$ states, differences with experiment are 
0.1--0.6~cm$^{-1}$ in Na, 2--4~cm$^{-1}$ in K, 1--7~cm$^{-1}$ in Rb, 9--24~cm$^{-1}$ in Cs,
and 13--29~cm$^{-1}$ in Fr. The removal energies of $np$ states are expected to be in better
agreement with experiment because of the smaller size of correlation corrections. 
We make predictions of $9p_{1/2}$ and $10p_{1/2}$ energies in Fr (where there are no
experimental values) in the last row
of Table \ref{tab2}. These predictions are based on the comparison of SD energies with experimental 
energies for other $np$ states in Cs and Fr. We expect our predictions to be accurate to 
about 5~cm$^{-1}$. Experimental energies for all states, except the $np$ states of Na, are larger than
theoretical values; in other words, correlation corrections are generally underestimated in the SD 
approximation.

The SD energies are compared with the relativistic CC calculations 
from Ref.~\cite{EKI} 
and with MBPT calculations from \cite{Fr} in Table~\ref{tab3}. 
The CC calculations agree better with experiment for $ns$ states 
except for the case of Na, where
the CC energy differs from experiment by about 100~cm$^{-1}$ for the $3s$ ground state. 
For the $np$ states, the present calculations are in better agreement with experiment than
the CC calculations, especially for the $6p_{1/2}$ state of Rb, and the $7p_{1/2}$ state of Cs.

The fine-structure intervals $np_{3/2}-np_{1/2}$ are compared
with experiment and with relativistic CC calculations
\cite{EKI} in Table \ref{tab4}.  Predictions for the fine-structure intervals
of the $8p$ and $9p$
states in Fr, based on comparisons of other intervals in Cs and Fr, are also
given in the table.    
The theoretical fine-structure intervals are seen to be 
in uniformly excellent agreement with experiment. 
  
In summary, the relativistic SD approximation 
gives accurate values for $ns$ removal energies in alkali-metal atoms, the
agreement with experiment being better for lighter elements.
Removal energies for $np$ states and $np_{3/2}-np_{1/2}$
fine-structure intervals are also in excellent agreement with experiment.  

\subsection{Electric-dipole matrix elements}

Electric-dipole matrix elements for $n'p_{1/2}-ns$ and $n'p_{3/2}-ns$ transitions
are evaluated in the SD approximation using the formalism laid out in \cite{liao}. 
In brief, the one-particle matrix element $Z$ is represented in second 
quantization as
\begin{equation}
Z = \sum_{ij} z_{ij} \, a_i^\dagger a_j  , 
\end{equation}
where $z_{ij}$ is the matrix element of the dipole operator $z$ between single-particle orbitals.
In the SD approximation, matrix elements of $Z$ are obtained by substituting SD wave function
from Eq.~(\ref{eqf}) into the matrix element $\langle \Psi_w | Z | \Psi_v \rangle$.
Correcting for normalization, one obtains:
\begin{equation}
\langle \Psi_w^{\rm SD} | Z | \Psi_v^{\rm SD}\rangle  = \delta_{wv} Z_{\rm core} + 
\frac{Z_{\rm val}^{\rm SD}}{\left[ (1+\delta N_{w}^{\rm SD} ) (1+\delta N_{v}^{\rm SD} ) \right]^{1/2}}, 
\end{equation}
where the first term contributes for scalar operators only.
The term $Z_{\rm val}^{\rm SD}$ is the sum 
\begin{equation}
 Z_{\rm val}^{\rm SD} = z_{wv}+z^{(a)}_{wv}+\cdots +  z^{(t)}_{wv},
\label{zval} 
\end{equation}
where $z_{wv}$ is the DHF matrix element and the remaining 20 terms are 
linear or quadratic functions of the single- and double-excitation coefficients 
$\rho_{ma}$, $\rho_{mv}$,  $\rho_{mnab}$ and $\rho_{mnva}$. Expressions for 
the terms $z^{(i)}_{wv}$ and the normalization constant $\delta N_{v}^{\rm SD}$ are given in Ref.\cite{liao}.
 
Matrix elements for $n'p_{1/2}-ns$ and $n'p_{3/2}-ns$ transitions 
with $n'=N \cdots N+3$ and $n=N,\; N+1$, where  
$N$ is the principal quantum number of the ground state, are calculated
using this method. 
The resulting  matrix elements are subsequently used to evaluate polarizabilities.
In Fr, electric-dipole matrix elements of $n'p-9s$ transitions are also calculated
to provide additional data for this least studied alkali-metal atom.

In Table~\ref{tab6}, we compare SD matrix elements for the principal
$Np_{1/2}-Ns$ and $Np_{3/2}-Ns$ transitions in Na, K, Rb, Cs, and Fr with the
high-precision experimental
results from \cite{V,S,nT}.  The differences between the present SD calculations 
and experiment range from 0.1\% in Na
to 0.5--0.8\% in Fr.
The SD results for the principal transitions are in all cases 
in better agreement with experiments
than the third-order MBPT values from \cite{MBPT} because of the more complete 
treatment of higher-order corrections.
In Cs, which has been extensively studied during the past fifteen years, 
all-order results from Refs.~\cite{D92} and \cite{CsDF}
are also available.  Comparison of our results with these theoretical calculations 
will be given below. 
Reduced matrix elements for transitions from all $n'p_{1/2}$ and $n'p_{3/2}$
states to  $Ns$ and $(N+1)s$ states of Na, K, and Rb 
are given in Table~\ref{tab7}. 
These matrix elements are used later to evaluate polarizabilities. 
Except for the principal transitions, no high-precision experimental 
values are available for these matrix elements. 

It is possible to include effects of triple excitations indirectly 
by using valence single- and double-excitation coefficients 
$\rho_{mv}$ and $\rho_{mnva}$ modified as explained previously to include triples partially.
Equation (\ref{zval}) itself is not modified in this procedure, thus,
effects of the triples are included only indirectly.
We use this procedure to obtain SDpT values for hyperfine constants. 
We found that including triples indirectly does not improve the agreement with experiment
for matrix elements of principal transitions, except for Na; for transitions other 
than the principal ones
the accuracy improves  slightly. 

To improve the accuracy of the matrix elements further,
one must include triple excitations {\em explicitly} in the matrix elements, i.e.\
calculate matrix elements in Eq.~(\ref{zval}) using the SDT wave function given in 
 Eq.~(\ref{SDT}). As a result,
expressions for $Z_{\rm val}$ will be modified:
\begin{equation}
 Z_{\rm val}^{\rm SDT} = Z_{\rm val}^{\rm SD}+ [{\rm triples} ],
\label{ztval} 
\end{equation}
where $[{\rm triples} ]$ are terms containing triple excitation coefficients $\rho_{mnrvab}$
and $\rho_{mnrabc}$.

It is possible to estimate the contribution of some omitted higher-order terms.
The dominant correlation corrections to most of the transition matrix elements are 
from the Brueckner-orbital (BO)
terms defined in \cite{Na} and discussed in Refs.~\cite{csao} and \cite{MBPT}.
To estimate the effect of omitted  higher-order corrections 
to the BO terms,  we scaled the single-excitation coefficients $\rho_{mv}$, as described
in Ref.~\cite{D92}: the coefficients 
were multiplied by the ratios of the experimental to theoretical correlation energies.
In Table~\ref{tabCS},  scaled results for Cs matrix elements are compared with our SD data,
with the all-order calculations of Refs.~\cite{D92,CsDF}, 
and with experiment.
The experimental data for $6p-6s$ transitions in this table are from the 
most recent measurement \cite{nT}. 
For the other transitions, with the exception of $7p-7s$, the experimental data compiled 
in Ref.\cite{CsDF} are used.
Matrix elements for $7p_{1/2}-7s$ and $7p_{3/2}-7s$ transitions can be determined 
accurately from a recent high-precision measurement of the Stark shift \cite{BW2}.  Values
determined in this way (described more completely in the following section) 
are listed instead of experimental data 
for these two transitions since no accurate experimental values are
available.  
We also list matrix elements from Ref.~\cite{10}, where experimental and theoretical data were compiled
to provide ``best values''. 
As we can see from the Table~\ref{tabCS} our {\it ab-initio} SD calculations provide accurate 
values for all of the matrix elements with the exception of $np-6s$. 
For $np-6s$ transitions, omitted higher-order corrections are very large, but can be estimated
using the scaling procedure described above.
Results from Refs.~\cite{D92} and \cite{CsDF} were obtained using similar scaling procedures,
 however, the relative importance of scaling is different in each case owing
to the different treatment of correlation corrections.  
In Ref.\cite{D92}, scaling gave small (0.2-0.4\%) contributions  for all  transitions \cite{csao},
while in Ref.\cite{CsDF} scaling led to 5.5\% and 4\% changes in $7p_{1/2}-6s$ and $7p_{3/2}-6s$ 
matrix elements and 0.1\% to 0.7\% changes in the others.
For our SD calculations, scaling changes matrix elements for $7p_{1/2}-6s$ and $7p_{3/2}-6s$ 
transitions by 6\% and 4\% respectively, and results for all other transitions by 0.5-1.2\%.
Our scaled matrix elements in Cs are in excellent agreement with 
other accurate theoretical results and with experimental values for transitions other than the 
principal transition. 
For the principal transition, the present scaled values are in poorer agreement with experiment, 
since scaling does not account for missing fourth- and higher-order RPA terms \cite{Na}  
that contribute significantly in this case. 
For other transitions, scaling of the SD results substantially increases the accuracy,
allowing us to make reasonable predictions for the corresponding transitions in Fr where no 
experimental results are available, and to estimate the accuracy of Fr polarizability 
calculations.

In Table~\ref{tabFr}, we compare our  results for $n's-np_{1/2}$ and  $n's-np_{3/2}$, 
matrix elements in francium with theoretical 
calculations from Refs.~\cite{Fr,MBPT,MVS}, and with experiment
\cite{S}. 
As for other alkali-metal atoms, the present all-order
results agree better with experiment than the MBPT results from Ref.~\cite{MBPT}. 
The results from the all-order calculations of \cite{Fr} are shown in column (a) of 
Table~\ref{tabFr} and the predictions
from \cite{Fr} are shown in column (b).  Our SD results 
for the $8s-7p$ transitions are between the values shown in columns (a) and (b), while SD data for
$8s-8p$ transitions are very close to values from (b). 
The only transition for which there is a large discrepancy between the SD values and those
from Ref.~\cite{Fr} is $7s-8p_{1/2}$.
As previously noted, there is a large contribution to this matrix element from triple excitations
that can be estimated by scaling.  
The scaled SD matrix element for this transition, listed in second column of
Table \ref{tabFr} is in
much better agreement with Ref.\cite{Fr}; it differs by 5\% from the result (a) and 
by 2\% from the prediction (b).
This transition is particularly hard to calculate since 
the total correlation correction is about
40\% (about twice as large as for the $7s-7p_{1/2}$ transition), and
a more accurate treatment of higher-order corrections is required.
Our SD result for the $7s-8p_{3/2}$ transition
agrees with Ref.~\cite {Fr} to 1\%. 
Recently, a large number of $n'p-ns$ matrix elements in Fr were evaluated using 
a semi-empirical model potential method \cite{MVS}.
These semi-empirical values agree with the {\it ab-initio} SD calculations
to better than 1\% with the exceptions of the
$7s-8p$ and $7s-9p$ transitions,
where contributions from correlation corrections are very large. 
The scaled SD data, which are more accurate for these four transitions, 
are in good agreement with \cite{MVS}.

In conclusion, the all-order SD method gives accurate data for a wide range
of $n'p-ns$ matrix elements for all alkali-metal atoms with exception of some 
transitions, such as $7s-8p$ in Fr, which have small dipole matrix elements and large 
correlation corrections. The accuracy  for such transitions
is significantly improved by scaling single excitation coefficients.
To achieve higher precision for electric-dipole matrix elements and to 
improve the accuracy of other matrix elements in Cs and Fr, a more complete treatment of 
triple excitations is necessary. 

\subsection{Static polarizabilities}
As mentioned in the introduction, SD matrix elements and energies were 
used to calculate static polarizabilities, Van der Walls coefficients, and atom-wall interaction coefficients 
of alkali-metal atoms in \cite{new}.
We discuss the calculation of the static polarizabilities
in more detail here.
The polarizabilities of the ground states of alkali-metal atoms
are given by the sum of two terms,
$\alpha = \alpha_v + \alpha_a$ where 
$\alpha_v$ is the contribution from valence excited states and $\alpha_a$ 
is the contribution from core excited (autoionizing) states. The contribution
of the autoionizing states can be well approximated by $\alpha_c$,
the polarizability of the ionic core. We write $\alpha_a = \alpha_c + \alpha_{cv}$, 
where $\alpha_{cv}$
is a counter term compensating for Pauli-principle violating excitations
from the core to the valence shell.
For an alkali atom in its $Ns$ ground state, these contributions are given by,
\begin{eqnarray}
\alpha_v &=&  \frac{1}{3}\sum_{n'} \left( 
  \frac{|\langle N s || z || n'p_{1/2} \rangle|^2} {E_{n'p_{1/2}}- E_{Ns}}
+ \frac{|\langle N s || z || n'p_{3/2} \rangle|^2} {E_{n'p_{3/2}}-E_{Ns}}\right)
\label{eqp1}, \\
\alpha_c &=& \frac{2}{3} 
\sum_{m a}  \frac{|\langle a || z || m \rangle|^2} {E_{m}-E_{a}}\label{eqp2}, \\
\alpha_{vc} &=& \frac{1}{3} \sum_a 
\frac{|\langle a || z || Ns \rangle|^2} {E_a-E_{Ns}}.
\label{eqp3}
\end{eqnarray}
The expressions for $\alpha_c$ and $\alpha_{vc}$ above are written in the single-particle
approximation. 

The dominant term is the valence contribution $\alpha_v$. This
term is evaluated by summing over the first few values of $n'$ in Eq.~(\ref{eqp1}) 
explicitly and approximating the remainder.
Thus, $\alpha_v= \alpha_v^{\rm main} + \alpha_v^{\rm tail}$.
In  the term $\alpha_v^{\rm main}$, we included $n'p$ states with 
$n'$=3--7 for Na, $n'$=4--7 for K, $n'$=5--8 for Rb,
$n'$=6--9 for Cs; and $n'$=7--10 for Fr. All matrix elements
were calculated using SD wave functions. These states account for more than 
99\% of $\alpha_v$; the small remainder $\alpha_v^{\rm tail}$ was evaluated 
in the DHF approximation and is expected to be accurate to better than 15\% for Na and
50\% for Fr.
The core polarizability $\alpha_c$  which contributes less than 10\% of the total
in all cases was calculated using the relativistic random-phase approximation (RRPA).
Values of $\alpha_c$ for Na$^+$, K$^+$, Rb$^+$, and Cs$^+$ were taken from
Ref.~\cite{RPA} and the RRPA value for Fr$^+$ was obtained in a separate 
calculation \cite{new}. The resulting values of $\alpha_c$ are expected to be accurate to 
better than 5\% based on
comparisons with recommended values from Miller and Bederson \cite{Bederson}. 
The much smaller valence-core contributions $\alpha_{vc}$ were
evaluated using DHF wave functions.

A breakdown of contributions to ground-state polarizabilities is 
given in Table~\ref{tabp}, together with a comparison with recommended 
values from \cite{new} and experiment \cite{prich,molof,hall}. 
In this table and in the paragraphs below, values of the polarizabilities are given in 
atomic units ($a_0^3$).
The SD results for Na, K, Rb, and Cs
are in excellent agreement with the values recommended in Ref.~\cite{new}
which were obtained using high-precision experimental matrix elements for
the principal transitions and experimental energies. In the case of Fr,
the difference is 1\%; however, the accuracy of the recommended value is
0.75\%. The difference in Fr is in part
due to the lower accuracy of the SD dipole matrix elements for the
principal transition compared to the accuracy of these matrix elements for other
alkalis.  

Stark-induced scalar and vector polarizabilities
$\alpha_S$ and $\beta_S$ for transitions from $Ns$ to the $(N+1)s$ states
were also calculated. The vector polarizability $\beta_S$ is important for 
the interpretation of PNC experiments \cite{Wieman}.
In addition, we evaluated differences $\Delta \alpha$ between 
polarizabilities of the $(N+1)s$ states and the $Ns$ ground states. 
Formulas for $\alpha_S$ and $\beta_S$ are given in \cite{D92}. 
Cesium is the only alkali-metal atom for which experimental data 
are available for all three of these parameters.
The present calculations provide useful reference data for the 
lighter alkali-metal atoms and for Fr. 

Contributions to $\alpha_S$ and $\beta_S$ are listed in
Table \ref{beta} together with comparisons with experiment 
and with semi-empirical calculations from Ref.~\cite{10}. 
The core contributions vanish for the Stark polarizabilities
but the core-valence contributions $\alpha_{vc}$ and $\beta_{vc}$ do not.
The terms $\alpha_S^{\rm tail}$, $\alpha_{vc}$, $\beta_S^{\rm tail}$ and
$\beta_{vc}$ were evaluated in the DHF approximation,
which is sufficient since these terms give small fractions of the totals.
The data in the rows labeled $\alpha_{S}^{SD}$ and $\beta_{S}^{SD}$ were obtained using SD
data for energies and matrix elements. 
The SD value for the scalar transition polarizability $\alpha_S$ in Cs differs from the
experimental value by 1.5\%. 
As we see from Table~\ref{beta}, $\beta_S$ is 
very small for Na but increase rapidly with $Z$. Our value of 26.87 for Cs
is in good agreement with the latest experimental value 
$\beta_S=27.024(43)_{\rm{expt}}(67)_{\rm{theory}}$ from Ref.~\cite{BW1}.
The vector polarizability $\beta_S$ is especially difficult to calculate precisely, since 
$np_{1/2}$ and $np_{3/2}$ terms contribute with opposite sign. 
For example,
the $6p_{1/2}$ contribution is -154.90 and the $6p_{3/2}$ contribution is 171.74.
 As a result, even small
uncertainties in the values of matrix elements can lead to large errors.
The principal uncertainties in $\beta_S$ are from $7s-6p$ and $7p-6s$ matrix elements. 
It should be noted
that it is sufficient to accurately know the ratio of 
($np_{3/2}-n^{\prime}s$)/($np_{1/2}-n^{\prime}s$) matrix elements
to significantly reduce the error. 

To estimate the accuracy of the SD value
and to provide recommended values for scalar and vector transition polarizabilities in Cs and Fr
we also calculate $\alpha_S$ and $\beta_S$ using experimental energies and
matrix elements for the principal transitions and scaled SD matrix elements for the other
transitions listed in Table \ref{tabCS}.  This semi-empirical calculation leads to
the recommended values in Table~\ref{beta} with the exception of the value of 
$\beta_S$ in Cs, which is listed in the separate row. The accuracy of the value of $\alpha_S$ 
is calculated assuming independent 
uncertainties in all matrix elements, where the uncertainties are based on the
comparisons with experiment.
The main contribution to the error in $\alpha_S$
comes from uncertainties in the $7p_{3/2}-6s$ matrix element,
which is accurate to 2\%.
The contribution of other uncertainties is insignificant.
The resulting value of $\alpha_S$ is in excellent agreement with the experimental value.
To estimate the accuracy of vector transition polarizability, we calculate $\beta_S$ 
using our recommended value of $\alpha_S$ and the high-precision experimental ratio 
$\alpha_S/\beta_S$=9.905(11) \cite{Cho}. The resulting
value of $\beta_S$, which is listed as the recommended value of Table \ref{beta} 
is 27.11(22) with error coming dominantly from the calculation of 
$\alpha_S$. As we see, this value is consistent with our direct calculation 
of $\beta_S$=27.16.
Further improvement in the accuracy of values of scalar and vector polarizability
will be possible when an accurate experimental value of the $7p_{3/2}-7s$ matrix element is obtained.
Our recommended values of $\alpha_S$ and $\beta_S$ in Cs are in excellent agreement
with values obtained by  Dzuba, Flambaum, and Sushkov \cite{10}. 
Uncertainties in the values of  $\alpha_S$ and $\beta_S$ in Ref.~\cite{10} are lower than 
the uncertainties of our recommended values
owing to the fact that a 0.7\% uncertainty
to the experimental value of $7p_{3/2}-6s$ matrix element is assigned in Ref.~\cite{10}.

We also carried out calculations of $\alpha_S$ and $\beta_S$ in Fr using both methods described above.
The results from the rows labeled ``Recomm.''  are obtained by using experimental values
of energies and principal transition matrix elements together with scaled SD data from Table~\ref{tabFr}. 
The SD results $\alpha_{S}^{SD}$ and $\beta_{S}^{SD}$ agree with our recommended values within 0.8\% 
for $\alpha_S$ and 1.4\% $\beta_S$. As in the case of Cs, the uncertainty in the value of
$\alpha_S$ is dominated by assuming that errors in all the transitions are independent.
The uncertainties are obtained by the uncertainty of the $8p_{3/2}-7s$ matrix element, which is 2\% based 
on comparison with Cs data. 
The final uncertainty in $\alpha_S$ in Fr is 1\%; the  
uncertainty in $\beta_S$ is also  1\% based on a comparison with Cs. 

Table \ref{alphad} gives the contribution to $\Delta \alpha$, the difference between
the static polarizabilities of the $(N+1)s$ states and the $Ns$ ground states
of alkali-metal atoms. 
The SD value $\Delta\alpha^{SD}$ of the scalar transition polarizability for Cs differs from the
recent experimental result 5837(6) \cite{BW2}  by 0.4\% and agrees within the error limits with 
the theoretical result 5833(80) from Ref.\cite{D92}. As noted previously,
the experimental value of $\Delta \alpha$ can be used to derive $7p-7s$ matrix elements to high accuracy, 
since $\Delta \alpha$ depends almost entirely on the values of these matrix elements.
The values of $7p_{1/2}-7s$ and $7p_{3/2}-7s$  matrix elements were varied to yield
experimental value of $\Delta \alpha$ within experimental precision. Ratio of these
matrix elements $D(7p_{3/2}-7s)/D(7p_{1/2}-7s)$ is taken to be 1.3892(3) based on the theoretical
calculations.
Experimental data were used for $6s-6p$, $6s-7p$, and $7s-6p$ matrix elements
and theoretical values were used for all others. The results
are $D(7p_{1/2}-7s)$=10.308(5) and $D(7p_{3/2}-7s)$=14.320(7)
assuming uncertainty only in the experimental value of $\Delta \alpha$. The only other significant
uncertainty is from the 0.5\% error in the value of $6p-7s$ matrix elements (which 
results in a 0.1\% variation in the value of $\Delta \alpha$). The final results, accounting
for the uncertainties in all matrix elements  and in the experimental value of $\Delta \alpha$, are
$D(7p_{1/2}-7s)$=10.308(15) and $D(7p_{3/2}-7s)$=14.320(20).
We give a recommended value for $\Delta\alpha$ in Fr obtained in the same
way as recommended value for $\alpha_S$. The uncertainty in this value comes almost
entirely from the uncertainty in the $8p-8s$ matrix elements, which is taken to be 0.3\%.

\subsection{Hyperfine Constants}
Results of our calculations of magnetic-dipole hyperfine constants $A$ (MHz)
for $ns$, $np_{1/2}$ and $np_{3/2}$ states in Na, K, Rb and Cs 
are given in Table \ref{tab9} together with experimental values from 
\cite{Happer,Wijngaarden,Yei,Rafac,Tanwei}. 
The nuclear magnetic moments used in the
calculations are weighted averages of values taken from the tabulation by Raghavan \cite{R};
they are listed in Table~\ref{tab10}. The calculations include corrections for the finite size 
of the nuclear magnetic moment distribution, which is modeled as a uniformly magnetized ball. 
The magnetization radii $R_m$ are obtained using nuclear parameters given in Ref.~\cite{JS} 
and are also listed in Table~\ref{tab10}. 
The rows labeled DHF in Table~\ref{tab9} give results calculated in the 
lowest-order DHF approximation.
The all-order results, including triple contributions as described in section \ref{Method}, are
listed in the rows labeled SDpT.
As stated in the introduction, the SD method gives poor results 
for ground-state hyperfine constants in alkalis, except for Na. 
In fact, the SD result for the  $6s$ hyperfine constant in Cs, without 
corrections for triples, overestimates the experimental value by 7\%, 
which is worse than the corresponding third-order MBPT result. 
As can be seen, the SDpT values are generally in excellent
agreement with experiment for $ns$ and $np_{1/2}$ states.
For the ground state of Cs, the agreement with experiment  
improves  to 1\%  using SDpT wave functions. 
Differences between SDpT results and experiment  are greater than 1\% for 
$np_{3/2}$ states of Rb and Cs. 
Further improvements of the accuracy of the hyperfine constants  
will require a more complete treatment of triples. 

In the calculations described above, corrections due to the finite size of the nuclear 
magnetic moment distribution (FS) in Na, K, and Rb are very small and are included in zeroth-order only.
However, FS corrections to hyperfine constants are significant for Cs and Fr and are, therefore,
included to all orders.  The relative size of the FS contributions to the correlation correction
in $ns$ states in these cases is found to be the same as in the lowest-order DHF calculation.  
Breit corrections to the hyperfine constants are calculated in second order following
the method outlined in \cite{M1}.
These corrections are  negligible for Na and K, but grow rapidly from 0.1\% for $5s$ state of Rb to 0.5\% 
for the $7s$ state of Fr. 

The SDpT values of hyperfine constants $A$ for the $7s$, $7p_{1/2}$, $7p_{3/2}$,
$8s$, $8p_{1/2}$, and $8p_{3/2}$ states in $^{211}$Fr are given in Table \ref{AFr},
where comparisons are made with experimental \cite{8s9s,ISOLDE,newFr} and
other theoretical data \cite{Dz}. 
It should be noted that FS corrections contribute 2.5\% to the $7s$ hyperfine constant. 
The values of $7s$ and $7p_{1/2}$  hyperfine constants for $^{211}$Fr differ 
from experimental values by 1.4\% and 1.8\%, respectively;
however, the accuracy of the magnetic moment $\mu=4.00(8)\,\mu_N$ \cite{ISOLDE} is 2\%. 
It should be noted that the SDpT result for the $6s$ state of Cs 
underestimates the experimental hyperfine constant by 1\%
but the SDpT result for the $7s$ state of Fr overestimates the experiment value by 1.4\%. 
The relative contribution of correlation for the Fr $7s$ hyperfine constant 
is about the same as for the Cs $6s$ hyperfine constant. 
Possible reasons for the anomalous differences with experiment are uncertainties in
the Fr magnetic moment or magnetic moment distribution; 
a more precise value of the magnetic moment is required to draw  
conclusions about the accuracy of the correlation correction.
The value of $A$ for the $7p_{3/2}$ state in Fr differs from experiment by 
3.2\%; however, it is lower than the experimental value, unlike  values for $7s$ and $7p_{1/2}$ states. 
The main source of theoretical uncertainty for the $7p_{3/2}$ hyperfine constant is the correlation correction, 
as it is for the Cs $6p_{3/2}$ hyperfine constant. 
Our results are in good agreement with theoretical calculation of \cite{Dz}, where
the Fr hyperfine constants were calculated using MBPT.
It should be noted that our results include Breit correction and a more complete treatment of the 
correlation and, thus, are expected to provide more accurate 
results for Fr hyperfine constants. 

The dependence of the FS correction on the value of magnetization radius was investigated in lowest order. 
Values of $A(7s)$ for Fr obtained with magnetization radii $R_m$=6.5~fm and $R_m$=7.0~fm but with the same 
charge radius $C_{nuc}$=6.71~fm differ by 0.2\%.  The $7s$ hyperfine constants calculated with 
$C_{nuc}$=$R_m$=6.5~fm and $C_{nuc}$=$R_m$=7.0~fm differ by 0.5\% of the total value. 

\subsection{Conclusion}

We have presented a systematic study of properties of alkali-metal atoms
using relativistic single-double wave functions. 
These wave functions give accurate values of removal energies,
fine-structure intervals, electric-dipole matrix elements, 
and polarizabilities for alkali-metal atoms from Na to Fr. 
The SD wave functions, however, lead to hyperfine constants for heavier alkali-metal
atoms that differ substantially from precise measurements. To obtain accurate
values for hyperfine constants, it was  necessary to include triples (partially) 
in the wave function.
This was done using the SDpT wave functions described in Section~\ref{Method} and
leads to accurate values of hyperfine constants.
Energies and transition matrix elements in Na determined here agree with those 
from the earlier SD calculation of Ref.~\cite{Na}; similarly, the present  
energies and matrix elements in Cs are in close agreement with the SD calculations 
of Ref.~\cite{D92}.  The SD calculations for 
K, Rb, and Fr presented here are completely new.  

The theoretical SD ground-state removal energies differ from 
experiment by amounts ranging from 2~cm$^{-1}$ in Na to 114~cm$^{-1}$ in Fr, and 
the SD removal energies for $np$  states agree with experimental values 
better than 30 cm$^{-1}$ for all states considered.
The theoretical SD matrix elements for principal transitions 
agree with recent high-precision experiments to 0.1-0.5\%, with exception of the $7s-7p_{3/2}$  
transition in Fr where the difference is 0.8\%. 
The agreement with experiment is better for lighter systems because of the smaller size 
of the correlation corrections.
A large number of matrix elements, which were used to calculate polarizabilities,
are tabulated for all alkali metal atoms; these matrix elements should
provide useful reference data.
The SD approximation  gives excellent results for static polarizabilities and for Stark-induced
transition polarizabilities. 
Supplementing our theoretical calculations with experimental energies and 
experimental matrix elements for the two principal transitions,
allowed us to predict values of the Stark polarizabilities $\alpha_S$ and $\beta_S$ for
in Cs and Fr to high accuracy.
The predicted values for $\alpha_S$ and $\beta_S$ in Cs
are in excellent agreement with experimental values. 
Hyperfine constants, calculated using SDpT wave functions, are in excellent agreement with
experiment for $ns$ and $np_{1/2}$ states of alkali-metal atoms from Na to Cs. 
Differences between theoretical SDpT ground-state hyperfine constants 
and experiment ranges from 0.3\% in Na to 1.4\% in Fr. The contributions of 
Breit and FS corrections to the ground state hyperfine constant in Fr are found 
to be significant. A more precise experimental value for the Fr nuclear magnetic moment
is necessary to evaluate the accuracy of correlation correctionc 
to Fr hyperfine constants. 

The methods developed in this work will be used in the future to evaluate 
PNC amplitudes in Cs and Fr.

\acknowledgements
This work was supported in part by National Science Foundation
Grants Nos.\ PHY-95-13179 and PHY-99-70666.

%\twocolumn

%\onecolumn

%\twocolumn

\begin{table}
\caption{Zeroth- (or DHF), second- and third-order MBPT removal energies $E^{(k)}$ in cm$^{-1}$ 
 and energy differences 
$\Delta^{(k)} = E_{\rm expt}-E^{(k)}$.}
\begin{tabular}{crrrrrr} 
\multicolumn{1}{c}{} &
\multicolumn{2}{c}{K ($4s$)} &
\multicolumn{2}{c}{Rb ($5s$)} &
\multicolumn{2}{c}{Cs ($6s$)} \\
\multicolumn{1}{c}{$k$} &
\multicolumn{1}{c}{$E^{(k)}$} &
\multicolumn{1}{c}{$\Delta^{(k)}$} &
\multicolumn{1}{c}{$E^{(k)}$} &
\multicolumn{1}{c}{$\Delta^{(k)}$} &
\multicolumn{1}{c}{$E^{(k)}$} &
\multicolumn{1}{c}{$\Delta^{(k)}$}\\
\hline 
$0$            & 32370&  2640& 30571&  3120& 27954& 3453\\
$2$            & 35104&   -94& 33878&  -187& 31865& -458\\
$3$            & 34655&   355& 33200&   491& 30529&  878\\
$E_{\rm expt}$ & 35010&      & 33691&      & 31407\\
\end{tabular}
\label{tab1}
\end{table}

\begin{table}
\caption{Comparison of SD calculations of $ns$ and $np_{1/2}$ removal energies
with experimental energies from \protect\cite{NIST,Fr,8s9s} in units of 
cm$^{-1}$. \label{tab2}}
\begin{tabular}{lcccc} 
\multicolumn{1}{l}{Na} &
\multicolumn{1}{c}{$3s$} &
\multicolumn{1}{c}{$4s$} &
\multicolumn{1}{c}{$5s$} &
\multicolumn{1}{c}{$6s$}  \\ \cline{1-5}
Theory & 41447.3& 15708.8 & 8248.48 & 5076.68 \\
Expt.  & 41449.4& 15709.4 & 8248.76 & 5076.82 \\[0.1pc]
Na &
\multicolumn{1}{c}{$3p_{1/2}$} &
\multicolumn{1}{c}{$4p_{1/2}$} &
\multicolumn{1}{c}{$5p_{1/2}$} &
\multicolumn{1}{c}{$6p_{1/2}$} \\ \cline{1-5}
Theory & 24493.9&  11183.0 & 6409.31 & 4153.22 \\
Expt.  & 24493.3&  11182.4 & 6409.06 & 4153.12 \\[0.2pc]
\multicolumn{1}{l}{K} &
\multicolumn{1}{c}{$4s$} &
\multicolumn{1}{c}{$5s$} &
\multicolumn{1}{c}{$6s$} &
\multicolumn{1}{c}{$7s$\rule[0.5pc]{0pt}{0.5pc}} \\ \cline{1-5}
Theory    &   34962  &   13958 &  7548 & 4730 \\
Expt.     &   35010  &   13986 &  7559 & 4735 \\[0.1pc]
K &
\multicolumn{1}{c}{$4p_{1/2}$} &
\multicolumn{1}{c}{$5p_{1/2}$} &
\multicolumn{1}{c}{$6p_{1/2}$} &
\multicolumn{1}{c}{$7p_{1/2}$} \\ \cline{1-5}
Theory    &   22023 &   10304 & 6008 & 3938 \\
Expt.     &   22025 &   10308 & 6010 & 3940  \\[0.2pc]
\multicolumn{1}{l}{Rb} &
\multicolumn{1}{c}{$5s$} &
\multicolumn{1}{c}{$6s$} &
\multicolumn{1}{c}{$7s$} &
\multicolumn{1}{c}{$8s$\rule[0.5pc]{0pt}{0.5pc}} \\ \cline{1-5}
Theory     &  33649 &  13527 & 7365 & 4637 \\
Expt.      &  33691 &  13557 & 7380 & 4644 \\[0.1pc]
Rb &
\multicolumn{1}{c}{$5p_{1/2}$} &
\multicolumn{1}{c}{$6p_{1/2}$} &
\multicolumn{1}{c}{$7p_{1/2}$} &
\multicolumn{1}{c}{$8p_{1/2}$} \\\cline{1-5}
Theory   &   21111&    9969  & 5852 & 3854 \\
Expt.    &   21112&    9976  & 5856 & 3856  \\[0.2pc] 
\multicolumn{1}{l}{Cs} &
\multicolumn{1}{c}{$6s$} &
\multicolumn{1}{c}{$7s$} &
\multicolumn{1}{c}{$8s$} &
\multicolumn{1}{c}{$9s$\rule[0.5pc]{0pt}{0.5pc}}  \\ \cline{1-5}
 Theory   & 31262 & 12801 & 7060 &  4479\\
   Expt.  & 31407 & 12871 & 7089 &  4496\\[0.1pc]
Cs &
\multicolumn{1}{c}{$6p_{1/2}$} &
\multicolumn{1}{c}{$7p_{1/2}$} &
\multicolumn{1}{c}{$8p_{1/2}$} &
\multicolumn{1}{c}{$9p_{1/2}$} \\ \cline{1-5}
 Theory  & 20204 &  9621 &  5687 &  3760 \\
   Expt. & 20228 &  9641 &  5698 &  3769 \\[0.2pc]  
\multicolumn{1}{l}{Fr} &
\multicolumn{1}{c}{$7s$} &
\multicolumn{1}{c}{$8s$} &
\multicolumn{1}{c}{$9s$} &
\multicolumn{1}{c}{$10s$ \rule[0.5pc]{0pt}{0.5pc}}  \\ \cline{1-5}
Theory   & 32735 & 13051 & 7148 & 4522 \\
Expt.    & 32849 & 13106 & 7168 & 4538\\[0.2pc]
Fr &
\multicolumn{1}{c}{$7p_{1/2}$} &
\multicolumn{1}{c}{$8p_{1/2}$} &
\multicolumn{1}{c}{$9p_{1/2}$} &
\multicolumn{1}{c}{$10p_{1/2}$} \\\cline{1-5}
Theory   & 20583 & 9712& 5724& 3782 \\
Expt.    & 20612 & 9736& {\it 5738} \tablenotemark[1]& {\it 3795}\tablenotemark[1]\\
\end{tabular}
\tablenotetext[1] {Prediction based on SD calculations.}
\end{table}

\begin{table}
\caption{Comparison of SD calculations of $ns$ and $np_{1/2}$ removal energies
with the CC calculations from Ref.~\protect\cite{EKI} and 
many-body calculations from Ref.~\protect\cite{Fr} in 
units of cm$^{-1}$. }
\begin{tabular}{ccccc} 
\multicolumn{1}{c}{Na} &
\multicolumn{1}{c}{$3s$} &
\multicolumn{1}{c}{$4s$} &
\multicolumn{1}{c}{$3p_{1/2}$} &
\multicolumn{1}{c}{$4p_{1/2}$} \\ \hline
SD & 41447& 15709& 24494& 11183\\
CC & 41352& 15690& 24465& 11172\\[0.2pc]
\multicolumn{1}{c}{K\rule[0.5pc]{0pt}{0.5pc}} &
\multicolumn{1}{c}{$4s$} &
\multicolumn{1}{c}{$5s$} &
\multicolumn{1}{c}{$4p_{1/2}$} &
\multicolumn{1}{c}{$5p_{1/2}$}\\ \hline
SD & 34962& 13958& 22023& 10304\\
CC & 35028& 13983& 22016& 10306\\[0.2pc]
\multicolumn{1}{c}{Rb\rule[0.5pc]{0pt}{0.5pc}} &
\multicolumn{1}{c}{$5s$} &
\multicolumn{1}{c}{$6s$} &
\multicolumn{1}{c}{$5p_{1/2}$}&
\multicolumn{1}{c}{$6p_{1/2}$}\\ \hline
SD & 33649& 13527& 21111& 9969\\
CC & 33721& 13564& 21117& 9857\\[0.2pc]
\multicolumn{1}{c}{Cs\rule[0.5pc]{0pt}{0.5pc}} &
\multicolumn{1}{c}{$6s$}&
\multicolumn{1}{c}{$7s$}&
\multicolumn{1}{c}{$6p_{1/2}$}&
\multicolumn{1}{c}{$7p_{1/2}$}\\ \hline
SD & 31262& 12801& 20204& 9621\\
CC & 31443& 12876& 20217& 9549\\[0.2pc]  
\multicolumn{1}{c}{Fr\rule[0.5pc]{0pt}{0.5pc}} &
\multicolumn{1}{c}{$7s$}&
\multicolumn{1}{c}{$8s$}&
\multicolumn{1}{c}{$7p_{1/2}$}&
\multicolumn{1}{c}{$8p_{1/2}$}\\ \hline
              SD &  32735& 13051& 20583 & 9712 \\
              CC &  32839& 13112& 20574 & 9736 \\
\protect\cite{Fr}&  32762& 13082& 20654&  9742\\
\end{tabular}
\label{tab3}
\end{table}

\begin{table}
\caption{Comparison of SD fine-structure intervals in Na, K, Rb, Cs, and Fr 
with experiment and with theoretical CC values from \protect\cite{EKI}. 
Units: cm$^{-1}$.
\label{tab4}} 
\begin{tabular}{lcrrr} 
\multicolumn{2}{c}{} &
\multicolumn{1}{c}{This work} &
\multicolumn{1}{c}{Expt.} & 
\multicolumn{1}{c}{\protect\cite{EKI}}  \\
\hline 
 Na & $3p_{3/2}-3p_{1/2}$ &    17.15&   17.20& 18.35\\
    & $4p_{3/2}-4p_{1/2}$ &    5.58 &    5.63&  5.99\\
    & $5p_{3/2}-5p_{1/2}$ &    2.46 &    2.52&      \\
    & $6p_{3/2}-6p_{1/2}$ &    1.26 &    1.25&     \\
    & $7p_{3/2}-7p_{1/2}$ &    0.75 &    0.74&     \\[0.1pc]
 K  & $4p_{3/2}-4p_{1/2}$ &    57.3 &   57.72& 59.45\\
    & $5p_{3/2}-5p_{1/2}$ &    18.5 &   18.8 & 19.3\\
    & $6p_{3/2}-6p_{1/2}$ &     8.5 &    8.4 &     \\
    & $7p_{3/2}-7p_{1/2}$ &     4.4 &    4.5 &     \\[0.1pc]
 Rb & $5p_{3/2}-5p_{1/2}$ &    236.5&   237.6& 240.3\\
    & $6p_{3/2}-6p_{1/2}$ &    76.5 &   77.5 &  87.7\\
    & $7p_{3/2}-7p_{1/2}$ &    34.8 &    35.1&     \\           
    & $8p_{3/2}-8p_{1/2}$ &    18.6 &    18.9&     \\[0.1pc]
 Cs & $6p_{3/2}-6p_{1/2}$ &    552.2&   554.1& 554.5\\
    & $7p_{3/2}-7p_{1/2}$ &    178.6&   181.0& 198.4\\
    & $8p_{3/2}-8p_{1/2}$ &     81.4&    82.6& \\
    & $9p_{3/2}-9p_{1/2}$ &     43.9&    44.7& \\[0.1pc]    
 Fr & $7p_{3/2}-7p_{1/2}$ &     1676&   1687 & 1670\\ 
    & $8p_{3/2}-8p_{1/2}$ &      536&   545  & 560\\ 
    & $9p_{3/2}-9p_{1/2}$ &      244&   250(3)\tablenotemark[1] & \\ 
    & $10p_{3/2}-10p_{1/2}$ &    132&   136(2)\tablenotemark[1]& \\ 
\end{tabular}
\tablenotetext[1] {Prediction based on SD calculations.}
\end{table}

%\onecolumn

\begin{table}
\caption{Comparison of SD calculations of reduced dipole matrix elements (a.u.)\  for 
the principal transitions in alkali-metal atoms with experimental values. 
 \label{tab6} }
\begin{tabular}{lllllllllllllll} 
\multicolumn{1}{c}{} &
\multicolumn{3}{c}{Na} &
\multicolumn{3}{c}{K} &
\multicolumn{3}{c}{Rb} &
\multicolumn{3}{c}{Cs} &
\multicolumn{2}{c}{Fr} \\
\multicolumn{1}{c}{} &
\multicolumn{1}{c}{$3p_{1/2}-3s$}&
\multicolumn{1}{c}{Ref.} &
\multicolumn{1}{c}{} &
\multicolumn{1}{c}{$4p_{1/2}-4s$} &
\multicolumn{1}{c}{Ref.} &
\multicolumn{1}{c}{} &
\multicolumn{1}{c}{$5p_{1/2}-5s$} &
\multicolumn{1}{c}{Ref.} &
\multicolumn{1}{c}{} &
\multicolumn{1}{c}{$6p_{1/2}-6s$} &
\multicolumn{1}{c}{Ref.} &
\multicolumn{1}{c}{} &
\multicolumn{1}{c}{$7p_{1/2}-7s$}&
\multicolumn{1}{c}{Ref.}\\
\hline
Present & 3.531      &   && 4.098   &    && 4.221    &   && 4.478     &     && 4.256 &\\
Expt.   & 3.5246(23) &\protect\cite{V} &
        & 4.102(5)   &\protect\cite{V} &
        & 4.231(3)   &\protect\cite{V} &
        & 4.4890(65) &\protect\cite{nT}&
        &4.277(8)    &\protect\cite{S} \\[1pc]     
\multicolumn{1}{l}{} &
\multicolumn{1}{l}{$3p_{3/2}-3s$}&
\multicolumn{1}{l}{Ref.} &
\multicolumn{1}{r}{} &
\multicolumn{1}{l}{$4p_{3/2}-4s$} &
\multicolumn{1}{l}{Ref.} &
\multicolumn{1}{r}{} &
\multicolumn{1}{l}{$5p_{3/2}-5s$} &
\multicolumn{1}{l}{Ref.} &
\multicolumn{1}{r}{} &
\multicolumn{1}{l}{$6p_{3/2}-6s$} &
\multicolumn{1}{l}{Ref.} &
\multicolumn{1}{r}{} &
\multicolumn{1}{l}{$7p_{3/2}-7s$} &
\multicolumn{1}{l}{Ref.} \\
\hline 
Present & 4.994     &   &&  5.794   &   &&   5.956    &   && 6.298     &    && 5.851 &\\
Expt.   & 4.9838(34)&\protect\cite{V} &
        &5.800(8)   &\protect\cite{V} &
        &   5.977(4)&\protect\cite{V} &
        & 6.3238(73)&\protect\cite{nT}&
        &5.898(15)  &\protect\cite{S} \\
\end{tabular}
\end{table}

%\twocolumn

\begin{table}
\caption{SD values of reduced dipole matrix elements (a.u.)\ 
in  Na, K, and Rb.
\label{tab7} }
\begin{tabular}{rrrrrrrr} 
\multicolumn{2}{c}{Na} &
\multicolumn{1}{c}{} &
\multicolumn{2}{c}{K} &
\multicolumn{1}{c}{} &
\multicolumn{2}{c}{Rb} \\
\hline
$3p_{1/2}-3s$&  3.531&&$4p_{1/2}-4s$ & 4.098 && $5p_{1/2}-5s$& 4.221  \\
$4p_{1/2}-3s$&  0.305&&$5p_{1/2}-4s$ & 0.275 && $6p_{1/2}-5s$& 0.333  \\
$5p_{1/2}-3s$&  0.107&&$6p_{1/2}-4s$ & 0.084 && $7p_{1/2}-5s$& 0.115  \\
$6p_{1/2}-3s$&  0.056&&$7p_{1/2}-4s$ & 0.039 && $8p_{1/2}-5s$& 0.059  \\[0.5pc]
$3p_{1/2}-4s$&  3.575&&$4p_{1/2}-5s$&  3.866&&  $5p_{1/2}-6s$& 4.119\\
$4p_{1/2}-4s$&  8.376&&$5p_{1/2}-5s$&  9.461&&  $6p_{1/2}-6s$& 9.684\\
$5p_{1/2}-4s$&  0.943&&$6p_{1/2}-5s$&  0.892&&  $7p_{1/2}-6s$&  0.999\\
$6p_{1/2}-4s$&  0.377&&$7p_{1/2}-5s$&  0.335&&  $5p_{1/2}-6s$&  0.393\\[0.5pc]
$3p_{3/2}-3s$&  4.994&&$4p_{3/2}-4s$ & 5.794&&  $5p_{3/2}-5s$  &  5.956\\
$4p_{3/2}-3s$&  0.435&&$5p_{3/2}-4s$ & 0.406&&  $6p_{3/2}-5s$  &  0.541\\
$5p_{3/2}-3s$&  0.154&&$6p_{3/2}-4s$ & 0.128&&  $7p_{3/2}-5s$  &  0.202\\
$6p_{3/2}-3s$&  0.081&&$7p_{3/2}-4s$ & 0.061&&  $8p_{3/2}-5s$  &  0.111\\[0.5pc]
$3p_{3/2}-4s$&  5.066&&$4p_{3/2}-5s$&   5.510&& $5p_{3/2}-6s$&   6.013\\
$4p_{3/2}-4s$& 11.840&&$5p_{3/2}-5s$&  13.358&& $5p_{3/2}-6s$& 13.592\\
$5p_{3/2}-4s$&  1.341&&$6p_{3/2}-5s$&   1.292&& $5p_{3/2}-6s$&  1.540\\
$6p_{3/2}-4s$&  0.537&&$7p_{3/2}-5s$&   0.491&& $5p_{3/2}-6s$&  0.628\\
\end{tabular}
\end{table}
%\onecolumn 

\begin{table}
\caption{Comparison of SD reduced dipole matrix elements (a.u.)\ for Cs
 with other theoretical values and with experiment.  \label{tabCS} }
\begin{tabular}{lrrrrll}
\multicolumn{1}{c}{Transition}&
\multicolumn{1}{c}{SD} &
\multicolumn{1}{c}{scaled} &
\multicolumn{1}{c}{Ref.\protect\cite{D92}} &
\multicolumn{1}{c}{Ref.\protect\cite{CsDF}} &
\multicolumn{1}{c}{Expt.}&
\multicolumn{1}{c}{Ref.\protect\cite{10}}\\
\hline
$6p_{1/2}-6s$&  4.482&   4.535&    4.510&    4.494&  4.4890(65)& \\
$6p_{3/2}-6s$&  6.304&   6.382&    6.347&    6.325&   6.3238(73)& \\[0.5pc]
$7p_{1/2}-6s$&  0.297&   0.279&    0.280&    0.275&  0.284(2)  & 0.2825(21)\\
$7p_{3/2}-6s$&  0.601&   0.576&    0.576&    0.583&   0.583(10) & 0.5820(44)\\[0.5pc]
$8p_{1/2}-6s$&  0.091&   0.081&    0.078&         &             &            \\
$8p_{1/2}-6s$&  0.232&   0.218&    0.214&         &             &            \\[0.5pc]
$6p_{1/2}-7s$&  4.196&   4.243&    4.236&    4.253&  4.233(22) &4.237(22)\\
$6p_{3/2}-7s$&  6.425&   6.479&    6.470&    6.507&   6.479(31) &6.472(31)\\[0.5pc]
$7p_{1/2}-7s$& 10.254&  10.310&   10.289&   10.288&  10.308(15)\tablenotemark[1]  &10.285(31)\\  
$7p_{3/2}-7s$& 14.238&  14.323&   14.293&   14.295&  14.320(20)\tablenotemark[1]  &14.286(43) \\
\end{tabular} 
\tablenotetext[1]{Predictions based on the experimental value of the Stark shift \protect\cite{BW2}.  }
\end{table}
%\onecolumn

\begin{table}
\caption{Comparison of SD reduced dipole matrix elements (a.u.)\ for Fr
 with other theoretical values and with experiment.  \label{tabFr} }
\begin{tabular}{rccccccc} 
\multicolumn{1}{c}{} &
\multicolumn{1}{c}{SD} &
\multicolumn{1}{c}{scaled} &
\multicolumn{1}{c}{Ref.\protect\cite{MVS}} &
\multicolumn{1}{c}{Ref.\protect\cite{Fr} \tablenotemark[1]} &
\multicolumn{1}{c}{Ref.\protect\cite{Fr} \tablenotemark[2]} &
\multicolumn{1}{c}{Ref.\protect\cite{MBPT}} &
\multicolumn{1}{c}{Expt.\protect\cite{S}} \\ \hline
   $7p_{1/2}-7s$&   4.256 & &      &  4.279   &   4.304  & 4.179  &    4.277\\
   $8p_{1/2}-7s$&   0.327 & 0.306 &   0.304&  0.291   &   0.301  &        &         \\
   $9p_{1/2}-7s$&   0.110 & 0.098 &   0.096&          &          &        &          \\
   $10_{1/2}-7s$&  0.055 & &   &          &          &        &          \\ [0.3pc]
   $7p_{3/2}-7s$&   5.851 & &     &  5.894   &  5.927   & 5.791  &    5.898\\
   $8p_{3/2}-7s$&   0.934 &  0.909 &   0.908&  0.924   &          &        &         \\
   $9p_{3/2}-7s$&   0.436 &  0.422 &   0.420&          &          &        &          \\
   $10p_{3/2}-7s$&  0.271 & &  &          &          &        &          \\[0.3pc]
   $7p_{1/2}-8s$&   4.184 & 4.237 &   4.230&  4.165   &    4.219 & 4.196  &          \\
   $8p_{1/2}-8s$&   10.02 & 10.10 &   10.06&  10.16   &    10.00 &        &         \\
   $9p_{1/2}-8s$&   0.985 &     &   0.977&          &         &        &          \\
   $10p_{1/2}-8s$&  0.380 & &   &          &         &        &          \\[0.3pc]
   $7p_{3/2}-8s$&   7.418 & 7.461 &   7.449&  7.384   &    7.470 & 7.472  &          \\
   $8p_{3/2}-8s$&   13.23 & 13.37 &   13.32&  13.45  &    13.26  &        &        \\
   $9p_{3/2}-8s$&   2.245 &       &   2.236&          &         &        &          \\
   $10p_{3/2}-8s$&  1.049 & &    &          &         &        &          \\[0.3pc]
   $7p_{1/2}-9s$&   1.016  & &  1.010& & & &\\
   $8p_{1/2}-9s$&   9.280 & &   9.342& & & &\\
   $9p_{1/2}-9s$&   17.39 & &   17.40& & & &\\
   $10p_{1/2}-9s$&  1.822 & &   & & & &\\[0.3pc]
   $7p_{3/2}-9s$&   1.393 & &   1.380& & & &\\
   $8p_{3/2}-9s$&   15.88 & &  15.92& & & &\\
   $9p_{3/2}-9s$&   22.59 & &  22.73& & & &\\
   $10p_{3/2}-9s$&  3.876 & &    & & & &
\end{tabular}
\tablenotetext[1]{Includes contributions from non-Brueckner diagrams extrapolated from Cs results.}
\tablenotetext[2]{Predictions given in \cite{Fr}.}
\end{table}

\begin{table}
\caption{Contributions to static polarizabilities (a.u.)\  of alkali-metal atoms and comparisons
with recommended values from \protect\cite{new}. \label{tabp} }
\begin{tabular}{lr@{.}lr@{.}lr@{.}lr@{.}lr@{}l}
\multicolumn{1}{l}{} &
\multicolumn{2}{c}{Na} &
\multicolumn{2}{c}{K} &
\multicolumn{2}{c}{Rb} &
\multicolumn{2}{c}{Cs} &
\multicolumn{2}{c}{Fr}\\ 
\hline
 $\alpha_v^{\rm main}$      & 162&06  & 284&70  & 308&43  & 383&8     & 294&.0\\
 $\alpha_v^{\rm tail}$      &   0&08  &   0&07  &   0&14  &   0&2     &   1&.4\\
 $\alpha_c$                 &   0&95  &   5&46  &   9&08  &  15&8     &  20&.4\\
 $\alpha_{vc}$              &  -0&02  &  -0&13  &  -0&26  &  -0&5     &  -0&.9\\
 $\alpha^{SD}$               & 163&07  & 290&10  & 317&39  & 399&3     & 314&.9\\
 Recomm. \protect\cite{new} & 162&6(3)& 290&2(8)& 318&6(6)& 399&9(1.9)& 317&.8(2.4) \\
Expt. &                       162&7(8)\tablenotemark[1]& 293&6(6.1)\tablenotemark[2]& 319&9(6.1)\tablenotemark[2]& 403&6(8.1)\tablenotemark[2]& 
\end{tabular}
\tablenotetext[1] {Ref.~\protect\cite{prich}.}
\tablenotetext[2] {Weighted average of experimental data from Refs.~\protect\cite{molof,hall}.}
\end{table}
\onecolumn
\begin{table}
\caption{Contributions to scalar and vector polarizabilities $\alpha_S$ and $\beta_S$ (a.u.)\ 
for  alkali-metal atoms. \label{beta}}
\begin{tabular}{lrrrr@{.}lr@{}l}
\multicolumn{1}{l}{} &
\multicolumn{1}{c}{Na} &
\multicolumn{1}{c}{K} &
\multicolumn{1}{c}{Rb} &
\multicolumn{2}{c}{Cs} &
\multicolumn{2}{c}{Fr}\\ 
\multicolumn{1}{l}{} &
\multicolumn{1}{c}{$3s-4s$} &
\multicolumn{1}{c}{$4s-5s$} &
\multicolumn{1}{c}{$5s-6s$} &
\multicolumn{2}{c}{$6s-7s$} &
\multicolumn{2}{c}{$7s-8s$}\\ 
\hline
 $\alpha_S^{\rm main}$   &     149.66 &  176.74 &   235.39 &    271&00&  374&.29\\  
 $\alpha_S^{\rm tail}$   &       0.32 &    0.28 &     0.68 &      0&87&    4&.22\\
 $\alpha_{vc}$           &      -0.01 &   -0.05 &    -0.11 &    -0&20 &   -0&.37\\
$\alpha_{S}^{SD}$        &      149.97&  176.97 &   235.96 &      271&67&  378&.14\\
Recomm.                  &            &         &          &    268&6(2.2)\tablenotemark[1]& 375&.3(3.6)\tablenotemark[1]\\
Dzuba {\it et al.} ~\protect\cite{10}& &         &          &    269&0(1.3)   &         \\
Expt.                    &            &         &          &    267&6(8)\tablenotemark[2] &  &     \\  [0.5pc]
 $\beta_S^{\rm main}$    &       0.35 &    1.95 &     9.18 &    26&77 &   72&.63\\ 
 $\beta_S^{\rm tail}$    &       0.00 &    0.00 &     0.04 &     0&10 &    0&.65\\  
 $\beta_{vc}$            &       0.00 &    0.00 &     0.00 &     0&00 &    0&.01\\ 
$\beta_{S}^{SD}$         &       0.35 &    1.95 &     9.22 &    26&87 &   73&.29\\
  &            &         &            &  27&16\tablenotemark[1]&   &     \\
Recomm. &            &         &          &    27&11(22)\tablenotemark[3] &   74&.3(7)\tablenotemark[1]\\
Dzuba {\it et al.} ~\protect\cite{10} &  &       &           &   27&15(13)   &         \\
Expt.                    &            &           &        &  27&02(8)\tablenotemark[4]&& \\
\end{tabular}
\tablenotetext[1] {Values obtained by using experimental values of energies and
matrix elements for the principal transitions and scaled SD data for the 8 other
transitions listed in Table \ref{tabCS} for Cs and Table \ref{tabFr} for Fr.}
\tablenotetext[2] {Value obtained by combining the measurement
of $\beta_S$~\protect\cite{BW1} with the accurately measured ratio 
$\alpha/\beta$ from Ref.~\protect\cite{Cho}.}
\tablenotetext[3] {Value obtained by using our recommended value of $\alpha$ and
the experimental $\alpha/\beta$ ratio from Ref.~\protect\cite{Cho}.}
\tablenotetext[4] {Ref.~\protect\cite{BW1}.}
\end{table}

\begin{table}
\caption{Contributions to the differences in static polarizabilities (a.u.)\ 
of $(N+1)s$ and the $Ns$ ground states of alkali-metal atoms. \label{alphad} }
\begin{tabular}{lrrrr@{}lr@{}l}
\multicolumn{1}{l}{} &
\multicolumn{1}{c}{Na} &
\multicolumn{1}{c}{K} &
\multicolumn{1}{c}{Rb} &
\multicolumn{2}{c}{Cs} &
\multicolumn{2}{c}{Fr}\\ 
\hline
$\Delta \alpha^{\rm main}$   &    2938.6  &  4673.7 &    4851.0&    5857&.1& 4419&.5 \\
$\Delta \alpha^{\rm tail}$   &       1.9  &     1.5 &       2.9&       3&.0&   11&.1 \\
$\Delta \alpha_{vc}$         &       0.0  &     0.1 &       0.2&       0&.4&    0&.8 \\
$\Delta \alpha^{SD}$           &    2940.5  &  4675.3 &    4854.1&    5860&.5& 4431&.4 \\
Recomm.     &           &         &          &          &&  4517&(26)\tablenotemark[1]\\
Expt.       &           &         &          &    5837&(6)\tablenotemark[2]& &     \\
\end{tabular}
\tablenotetext[1] {Value obtained using experimental energies and either experimental or
scaled SD matrix elements.}
\tablenotetext[2] {Ref.~\protect\cite{BW2}.}
\end{table}

%\twocolumn
\begin{table}
\caption{ Comparison of SDpT values of hyperfine constants $A$ (MHz) of 
$ns$, $np_{1/2}$, and $np_{3/2}$ states
of alkali-metal atoms with experiment. 
Experimental values are from Ref.~\protect\cite{Happer}, unless noted
otherwise. }
\begin{tabular}{lllll} 
\multicolumn{1}{c}{Na} &
\multicolumn{1}{l}{$3s$} &
\multicolumn{1}{l}{$4s$} &
\multicolumn{1}{l}{$5s$} &
\multicolumn{1}{l}{$6s$} \\
\hline
DHF  & 623.8 & 150.5 & 58.04 & 28.21\\
SDpT & 888.3 & 204.3 & 77.68 & 37.51\\
Expt.& 885.8 & 202(3)& 78(5)      &\\[0.1pc]
\multicolumn{1}{c}{Na} &                        
\multicolumn{1}{l}{$3p_{1/2}$} &                        
\multicolumn{1}{l}{$4p_{1/2}$} &                        
\multicolumn{1}{l}{$5p_{1/2}$} &                        
\multicolumn{1}{l}{$6p_{1/2}$} \\                       
\hline                                          
 DHF  & 63.4    & 21.0 &  9.3 &  4.9\\          
SDpT & 95.1    & 30.7 & 13.5 &  7.1\\          
Expt.& 94.44(13)\tablenotemark[1]  &      &       &    \\[0.1pc]         
\multicolumn{1}{c}{Na} &                        
\multicolumn{1}{l}{$3p_{3/2}$} &                        
\multicolumn{1}{l}{$4p_{3/2}$} &                        
\multicolumn{1}{l}{$5p_{3/2}$} &                        
\multicolumn{1}{l}{$6p_{3/2}$} \\                       
\hline                                          
DHF   & 12.6      & 4.16    & 1.85 &  0.98\\          
SDpT & 18.8      & 6.04    & 2.66 &  1.40\\          
Expt.& 18.534(15)\tablenotemark[2]   & 6.01(3) &       &    \\[0.3pc]   
\multicolumn{1}{c}{K} &
\multicolumn{1}{l}{$4s$} &
\multicolumn{1}{l}{$5s$} &
\multicolumn{1}{l}{$6s$} &
\multicolumn{1}{l}{$7s$} \\
\hline
DHF    & 146.8 & 38.85 &15.75 & 7.89  \\
SDpT  & 228.6 & 54.81 &21.61 &10.68 \\
Expt. & 230.85& 55.50(60)& 21.81(18)&10.85(15)\\[0.3pc]
\multicolumn{1}{c}{K} &                        
\multicolumn{1}{l}{$4p_{1/2}$} &                        
\multicolumn{1}{l}{$5p_{1/2}$} &                        
\multicolumn{1}{l}{$6p_{1/2}$} &                        
\multicolumn{1}{l}{$7p_{1/2}$} \\                       
\hline                                          
DHF  &   16.61  &   5.74 &    2.62 &    1.41\\
SDpT &  27.65  &   8.95 &    4.02 &    2.14\\
Expt.&  28.85(30) & 8.99(15) &         & \\[0.1pc]         
\multicolumn{1}{c}{K} &                        
\multicolumn{1}{l}{$4p_{3/2}$} &                        
\multicolumn{1}{l}{$5p_{3/2}$} &                        
\multicolumn{1}{l}{$6p_{3/2}$} &                        
\multicolumn{1}{l}{$7p_{3/2}$} \\                       
\hline                                          
DHF  &   3.23 &     1.11&    0.512&      0.276\\
SDpT &  5.99 &     1.93&    0.866&      0.462\\
Expt.&  6.09(4) &  1.97(1)  &0.866(8) &      \\[0.3pc]   
\multicolumn{1}{c}{Rb} &
\multicolumn{1}{l}{$5s$} &
\multicolumn{1}{l}{$6s$} &
\multicolumn{1}{l}{$7s$} &
\multicolumn{1}{l}{$8s$} \\
\hline
DHF  & 642.6  & 171.6 & 70.3& 35.5\\
SDpT & 1011.1 & 238.2 &94.3& 46.9\\
Expt.& 1011.9 & 239.3(1.2)&94.00(64)&45.5(2.0)\\[0.3pc]
\multicolumn{1}{c}{Rb} &                        
\multicolumn{1}{l}{$5p_{1/2}$} &                        
\multicolumn{1}{l}{$6p_{1/2}$} &                        
\multicolumn{1}{l}{$7p_{1/2}$} &                        
\multicolumn{1}{l}{$8p_{1/2}$} \\                       
\hline                                          
DHF  &    69.8 &   24.55 &  11.39  &   6.19\\
SDpT &  120.4  &  39.02  & 17.61 &    9.45\\
Expt.&  120.7(1)& 39.11(3)& 17.65(2) & \\[0.1pc]         
\multicolumn{1}{c}{Rb} &                        
\multicolumn{1}{l}{$5p_{3/2}$} &      \multicolumn{1}{l}{$6p_{3/2}$} &                        
\multicolumn{1}{l}{$7p_{3/2}$} &                        
\multicolumn{1}{l}{$8p_{3/2}$} \\                       
\hline                                          
DHF  &   12.4 &    4.37&     2.03&     1.11\\
SDpT &  24.5 &    7.98 &    3.61&     1.94\\ 
Expt.& 25.029(16)& 8.25(10) & 3.71(1)  &      \\[0.3pc]   
\multicolumn{1}{c}{Cs} &
\multicolumn{1}{l}{$6s$} &
\multicolumn{1}{l}{$7s$} &
\multicolumn{1}{l}{$8s$} &
\multicolumn{1}{l}{$9s$} \\
\hline
DHF   & 1425.2 & 391.6 &   163.5 & 83.6  \\
SDpT & 2278.5 & 540.6 &   217.1&  109.1 \\  
Expt.& 2298.2 &545.90(9)& 218.9(1.6)& 109.5(2.0)\\[0.1pc]
\multicolumn{1}{c}{Cs} &
\multicolumn{1}{l}{$6p_{1/2}$} &
\multicolumn{1}{l}{$7p_{1/2}$} &
\multicolumn{1}{l}{$8p_{1/2}$} &
\multicolumn{1}{l}{$9p_{1/2}$} \\
\hline
DHF  & 160.9    & 57.62 &  27.08 &    14.84\\
SDpT & 289.6    & 93.40 &  42.43 &    22.76\\       
Expt.& 291.89(8)\tablenotemark[3] & 94.35 & 42.97(10)& \\[0.1pc]
\multicolumn{1}{c}{Cs} &
\multicolumn{1}{l}{$6p_{3/2}$} &
\multicolumn{1}{l}{$7p_{3/2}$} &
\multicolumn{1}{l}{$8p_{3/2}$} &
\multicolumn{1}{l}{$9p_{3/2}$} \\
\hline
DHF  & 23.93     & 8.64       &   4.08   &   2.24\\
SDpT & 48.51      & 15.88      & 7.27   & 3.93    \\  
Expt.& 50.275(3)\tablenotemark[4] & 16.605(6) &7.626(5)   & 4.129(7)
\end{tabular}
\tablenotetext[1] {Ref.~\protect\cite{Wijngaarden}.}
\tablenotetext[2] {Ref.~\protect\cite{Yei}.}
\tablenotetext[3] {Ref.~\protect\cite{Rafac}.}
\tablenotetext[4] {Ref.~\protect\cite{Tanwei}.}
\label{tab9}
\end{table}

\begin{table}
\caption{Nucleon numbers A, nuclear spins $I$, 
magnetization radii $R_m$(fm) from \protect\cite{JS},
and magnetic moments $\mu_I$ in units of $\mu_N$ from \protect\cite{R} 
used in the preparation
of Table~\protect\ref{tab9}. \label{tab10}}
\begin{tabular}{lllll} 
\multicolumn{1}{c}{} &
\multicolumn{1}{c}{A} &
\multicolumn{1}{c}{$I$} &
\multicolumn{1}{c}{$R_m$} &
\multicolumn{1}{c}{$\mu_I$} \\
\hline
Na & 23  & 3/2 & 2.89& 2.2176 \\
K  & 39  & 3/2 & 3.61& 0.39149 \\
Rb & 85  & 5/2 & 4.87& 1.3534 \\
Cs & 133 & 7/2 & 5.67& 2.5826 \\
\end{tabular}
\end{table}

\begin{table}
\caption{Comparison of SDpT values of Fr hyperfine constants $A$ (MHz) with experiment
and other theory. $g_I$=0.888, $R_m$=6.71fm \label{AFr} }
\begin{tabular}{lllllll} 
\multicolumn{1}{c}{} &
\multicolumn{1}{l}{$7s$} &
\multicolumn{1}{l}{$7p_{1/2}$} &
\multicolumn{1}{l}{$7p_{3/2}$} &
\multicolumn{1}{l}{$8s$} &
\multicolumn{1}{l}{$8p_{1/2}$} &
\multicolumn{1}{l}{$8p_{3/2}$} \\
\hline
DHF    & 5785.7   &  622.7 &     49.30 & 1482.8 &  220.91&  18.03\\
SDpT   & 8833.0   & 1162.1 &     91.80 & 1923.3 &  362.91&  30.41\\
Expt.  & 8713.9(8)\tablenotemark[1]& 1142.0(3)\tablenotemark[2]&  94.9(3)\tablenotemark[1]&1912.5(1.3)\tablenotemark[3]    &   & \\
Ref.~\protect\cite{Fr}& 9018  &  1124   &    102.2 &   1970 &     363.6 &  35.2\\
\end{tabular}
\tablenotetext[1]{Ref.~\protect\cite{ISOLDE}.}
\tablenotetext[2] {Ref.~\protect\cite{newFr}. }
\tablenotetext[3] {Value obtained by rescaling experimental value for $^{210}$Fr 
1577.8(1.1) MHz from Ref.~\protect\cite{8s9s} using $\mu$(210)=4.40$\mu_N$ 
and $\mu$(211)=4.00$\mu_N$. The uncertainty includes
experimental uncertainty of $^{210}$Fr value 1577.8(1.1) MHz only.}
\end{table}

\begin{references}
\bibitem{liao}      S.A. Blundell, W.R. Johnson,  Z.W. Liu, and J. Sapirstein,  
                    Phys.\ Rev.\ A{\bf 40}, 2233 (1989).

\bibitem{liu}       Z.W. Liu, Ph.\ D. thesis, Notre Dame University, (1989).

\bibitem{csao}      S.A. Blundell, W.R. Johnson, and J. Sapirstein,
                    Phys.\ Rev.\ A{\bf 43}, 3497 (1991). 

\bibitem{Na}        M.S. Safronova, A. Derevianko, and W.R. Johnson, 
                    Phys.\ Rev.\ A{\bf 58}, 1016 (1998).

\bibitem{EKI}       E. Eliav, U. Kaldor, and Y. Ishikawa,
                    Phys.\ Rev.\ A{\bf 50}, 1121 (1994).

\bibitem{J}         K.M. Jones, P.S. Julienne, P.D. Lett, W.D. Phillips, E. Tiesinga, 
                    and C.J. Williams, 
                    Europhys.\ Lett.\ {\bf 35}, 85 (1996).

\bibitem{V}         U. Volz and H. Schmoranzer, 
                    Phys.\ Scr.\ T{\bf 65}, 48 (1996). 

\bibitem{W}         H. Wang,   P.L. Gould, W.C. Stwalley, 
                    J. Chem.\ Phys.\ {\bf 106}, 7899 (1997).

\bibitem{S}         J.E. Simsarian,  L.A. Orozco, G.D. Sprouse, and W.Z. Zhao, 
                    Phys.\ Rev.\ A{\bf 57}, 2448 (1998). 

\bibitem{oT}        R.J. Rafac, C.E. Tanner, A.E. Livingston, 
                    K.W. Kukla, H.G. Berry, and C.A. Kurtz, 
                    Phys.\ Rev.\ A{\bf 50}, 1976 (1994). 

\bibitem{Y}         L. Young, W.T. Hill III, S.J. Sibener, S.D. Price,
                    C.E. Tanner, C.E. Wieman, and S.R. Leone, 
                    Phys.\ Rev.\ A {\bf 50}, 2174 (1994).

\bibitem{nT}        R.J. Rafac, C.E. Tanner, A.E. Livingston, and H.G. Berry,
                    Phys.\ Rev.\ A (submitted for publication 1998).

\bibitem{new}       A. Derevianko, W.R. Johnson, M.S. Safronova, and J.F. Babb, 
                    Phys.\ Rev.\ Lett.\ {\bf 82}, 3589  (1999).

\bibitem{Wieman}    C. S. Wood, S. C. Bennett, D. Cho, B. P. Masterson, J. L. Roberts, 
                    C. E. Tanner, and C. E. Wieman, Science, {\bf 275}, 1759 (1997).

\bibitem{nopair}    G.E. Brown and D.G. Ravenhall, 
                    Proc.\ R.\ Soc.\ London, Ser.\ A {\bf 208}, 552 (1951).

\bibitem{31}        W.R. Johnson, S.A. Blundell, and J. Sapirstein, 
                    Phys.\ Rev.\ A{\bf 37}, 2764 (1988).     

\bibitem{NIST}      C.E. Moore, {\it Atomic Energy Levels}, Natl.\ Bur.\ 
                    Stand.\ Ref.\ Data Ser., Natl.\ Bur.\ Stand.\ (U.S.) 
                    Circ.\ No.\ 35 (U.S. GPO, Washington, D.C., 1971), Vols.\ I-III.
                    
\bibitem{Fr}        V.A. Dzuba, V.V. Flambaum, and O.P. Sushkov, 
                    Phys.\ Rev.\ A {\bf 51}, 3454 (1995).

\bibitem{8s9s}      J.E. Simsarian, W. Shi, L.A. Orozco, G.D. Sprouse, W.Z. Zhao,
                    Opt.\ Lett.\ {\bf 21} (1996), 1939; 
                    J.E. Simsarian, W.Z. Zhao, L.A. Orozco, G.D. Sprouse, 
                    Phys.\ Rev.\ A {\bf 59} (1999) 195.

\bibitem{MBPT}      W.R. Johnson, Z.W. Liu and J. Sapirstein, 
                    At.\ Data and Nucl.\ Data Tables {\bf 64}, 279 (1996); 
                    J. Sapirstein, 
                    Rev.\ Mod.\ Phys.\ {\bf 70}, 55 (1998).

\bibitem{D92}       S.A. Blundell, J. Sapirstein, and W.R. Johnson,  
                    Phys.\ Rev.\ D{\bf 45}, 1602 (1992).    
                    
\bibitem{CsDF}      V.A. Dzuba, V.V. Flambaum, and O.P. Sushkov, Phys. Lett. A {\bf 141}, 
	            147 (1989).

\bibitem{BW2}       S.C. Bennett, J.L. Roberts, and C.E. Wieman, 
                    Phys.\ Rev.\ A. {\bf 59}, R16 (1999).

\bibitem{10}        V.A. Dzuba, V.V. Flambaum, and O. P. Sushkov, Phys.\ Rev.\ A, 
                    {\bf 56} 4357 (1997).                 
  
\bibitem{MVS}       M. Marinescu, D. Vrinceanu, and H.R. Sadeghpour, 
                    Phys.\ Rev.\ A {\bf 58}, R4259 (1998).
    
\bibitem{RPA}       W.R. Johnson, D. Kolb, and K.-N.\ Huang, 
                    At.\ Data Nucl.\ Data Tables {\bf 28}, 333 (1983).
 
\bibitem{Bederson}  T.M. Miller and B. Bederson,
                    Adv.\ At.\ Mol.\ Phys.\ {\bf 13}, 1 (1977).

\bibitem{prich}     C.R. Ekstr\"{o}m, J. Schmiedmayer, M.S. Chapman, C.D. Hammond,
                    and D.E. Pritchard,
                    Phys.\ Rev.\ A{\bf 51}, 3883 (1995).

\bibitem{molof}     R.W. Molof, H.L. Schwartz, T.M. Miller, B. Bederson,
                    Phys.\ Rev.\ A {\bf 10}, 1131 (1974).
     
\bibitem{hall}      W.D. Hall and J.C. Zorn, Phys. Rev. A, {\bf 10}, 1141 (1974).

\bibitem{BW1}       S.C. Bennett and C.E. Wieman, 
                    Phys.\ Rev.\ Lett {\bf 82}, 2484 (1999).

\bibitem{Cho}       D. Cho, C.S. Wood, S.C. Bennett, J.L. Roberts, and C.E. Wieman,
                    Phys.\ Rev.\ A {\bf 55}, 1007 (1997).
\bibitem{Happer}    W. Happer in {\it Atomic Physics 4}, eds.\ G. zu Putlitz, E.W. Weber, 
                    and A. Winnacker, (Plenum Press, New York, 1974) pp.\ 651-682. 
		    
\bibitem{Wijngaarden} W.A. Wijngaarden and J. Li, Z.\ Phys.\ D {\bf 32}, 67 (1994).

\bibitem{Yei}       W. Yei, A. Sieradzan and M.D. Havey, Phys.\ Rev.\ A {\bf 48}, 1909 (1993) . 

\bibitem{Rafac}     R.J. Rafac, C.E. Tanner,
                    Phys.\ Rev.\ A {\bf 56}, 1027 (1997).
 
\bibitem{Tanwei}    C.E. Tanner and C. Wieman,
                    Phys.\ Rev.\ A {\bf 38}, 1616 (1988).
   
\bibitem{R}         P. Raghavan, 
                    At.\ Data Nucl.\ Data Tables {\bf 42}, 189 (1989).

\bibitem{JS}        W.R. Johnson and G. Soff, 
                    At.\ Data Nucl.\ Data Tables {\bf 33}, 405 (1985). 

\bibitem{M1}        I.M. Savukov, A. Derevianko, H.G. Berry, and W.R. Johnson, submitted
                    to Phys.\ Rev.\ Lett.\
    
\bibitem{ISOLDE}    C. Ekstr\"{o}m, L. Robertsson, A. Ros\'{e}n, and the ISOLDE collaboration, 
                    Phys.\ Scr.\ {\bf 34}, 624 (1986).

\bibitem{newFr}     J.S. Grossman, L.A. Orozco, M.R. Pearson, J.E. Simsarian, G.D. Sprouse,
                    and W.Z. Zhao, submitted to Phys.\ Rev.\ Lett.

\bibitem{Dz}        V.A. Dzuba, V.V. Flambaum, and O.P. Sushkov, J.\ Phys.\ B {\bf 17}, 1953 (1984).
\end{references}
\end{document}